\documentclass[journal,twoside]{IEEEtran}

\usepackage{amsmath,amssymb}
\usepackage{graphicx}
\graphicspath{{}}
\usepackage{dcolumn}
\usepackage{bm}
\usepackage{caption,subcaption}
\usepackage{ulem}
\usepackage[utf8]{inputenc}
\usepackage[T1]{fontenc}
\usepackage{mathptmx}
\usepackage[dvipsnames]{xcolor}
\renewcommand{\sout}[1]{}
\renewcommand{\uline}[1]{#1}

\begin{document}

\title{A Simple Coupled-Bloch-Mode Approach To Study Active Photonic Crystal Waveguides and Lasers}

\author{\thanks{Manuscript received xxx, 2019; revised YYY, 2019.}
Marco Saldutti, Paolo Bardella, Jesper Mørk and Mariangela Gioannini\thanks{M. Saldutti, P. Bardella and M. Gioannini are with Politecnico di Torino, Electronic and Telecommunication Department, Italy, \mbox{email:marco.saldutti@polito.it}}
\thanks{J. Mørk is with Technical University of Denmark, Photonics Engineering Department, Denmark, \mbox{email:jesm@fotonik.dtu.dk}}
}



\maketitle

\begin{abstract}
By applying a coupled-Bloch-mode approach, we have derived a simple expression for the transmission properties of photonic crystal (PhC) line-defect waveguides with a complex refractive index perturbation. We have provided physical insights on the coupling mechanism by analyzing the frequency dependence and relative strength of the coupling coefficients. We have shown the impact of the perturbation on the waveguide dispersion relation and how the gain-induced distributed feedback limits the maximum attainable slow-light enhancement of the gain itself. We have then applied our approach to analyze the threshold behaviour of various PhC laser cavities and proved the significant impact of coherent distributed feedback effects in these lasers. Importantly, our approach also reveals that a structure simply consisting of an active region with zero back reflections from the passive output waveguides can achieve lasing oscillation with reasonable threshold gain.    
\end{abstract}


\IEEEpeerreviewmaketitle

\section{Introduction}

\IEEEPARstart{P}{hotonic}-crystal (PhC) waveguides are made by creating a line-defect in a PhC slab. The defect introduces guided modes in the photonic band-gap of the crystal and allows for an efficient propagation of optical signals. In particular, a major advantage of PhC line-defect waveguides (LDWGs), as compared to conventional waveguides, is the possibility to exploit slow-light (SL) propagation \cite{Baba2008}. In the SL region of the waveguide dispersion relation, the group velocity is greatly reduced and ideally tends to zero at the band edge. As a consequence, in the SL region, a Bloch wave propagating within an active PhC waveguide experiences an effective gain per unit length which is greatly enhanced with respect to the modal gain coefficient of a conventional waveguide \cite{ek2014a}. This gain-enhancement allows for the realization of shorter devices and makes PhC waveguides ideal candidates for high-density photonic integrated circuits. PhC lasers have attracted large research as efficient light-sources in on-chip and chip-to-chip interconnections. They allow for scaling the active volume while maintaining a high cavity Q-factor, thus exhibiting low threshold current and operating energy \cite{MatsuoReview}. Different types of PhC lasers exist. A typical PhC laser consists of a so called $\mathrm{LN}$ cavity \cite{Okano2010}, that is formed by only omitting $N$ holes in a PhC slab. Initially, due to heat accumulating in the active region, room-temperature continuous-wave (RT-CW) operation was difficult to achieve in this type of PhC laser. In the last years, significant progress has been made and RT-CW operation has been demonstrated under optical pumping by using ultra-small cavities with embedded quantum dots (QDs) \cite{Nomura2006}. A major breakthrough, addressing the issues of high-speed direct modulation and thermal stability, has been the introduction of the so called lambda-scale-embedded active region PhC laser (LEAP laser), working under electrical \cite{Matsuo2012} pumping conditions. In this laser, the cavity is made of a buried heterostructure (BH) embedded in a PhC LDWG. Furthermore, differently from PhC lasers based on $\mathrm{LN}$ cavities \cite{Xue2015,Xue2016}, in LEAP lasers light can be emitted into in-plane waveguides. The output waveguide can be either shifted with respect to the active region \cite{MatsuoNatPhot2013,MatsuoJSTQE2013} or placed in-line with it \cite{Shinya}. Therefore, LEAP lasers are promising sources for photonic integrated circuits \cite{MatsuoReview}.
\IEEEpubidadjcol

Rigorous approaches to analyze PhC devices, such as FDTD \cite{Okano2010,Okano2009} or RCWA \cite{LalanneRCWA2005,LalanneRCWA2007,SongRCWA}, are time- and memory-consuming and cannot always provide an intuitive understanding of the physical phenomena into play. The aim of this paper is to present an alternative and simple approach for analyzing active PhC LDWGs and lasers based on this type of waveguides. Our approach is based on coupled-mode theory (CMT) \cite{Yariv1973,Marcuse_Book1974}, which has proved to be an effective tool to study and design lasers with periodic gain and/or refractive index perturbation, such as edge-emitting DFB lasers \cite{KogelnikShank}. More recently, CMT has also been used to analyze both passive \cite{Patterson2010,MichaelisPR} and active \cite{Chen2015} PhC waveguides. By following a simple CMT approach, we show that a weak perturbation of the gain and/or refractive index of a PhC waveguide causes a strong coupling between the forward- and backward-propagating Bloch modes of the unperturbed waveguide, with coupling coefficients strongly increasing as we move towards the band edge. We start from the formulation already presented in \cite{Chen2015}, where the Bloch modes of a passive, reference waveguide were used as a basis for the field expansion in a finite-length, active section; the presence of gain in this section was treated as a weak perturbation to the passive structure, coupling the otherwise independent, counter-propagating Bloch modes. However, the system of coupled propagation equations was solved numerically in \cite{Chen2015}, thus not providing important insight on the coupling mechanism. In this work, we push further the formulation of \cite{Chen2015} to generally take into account both a real and imaginary refractive index perturbation. Avoiding the numerical solution, we derive a simple, closed-form expression for the unit cell transmission matrix of an active PhC waveguide. Therefore, consistently with the rigorous, non-perturbative approach of \cite{Gric2012}, we show that, in the presence of gain, the group index does not diverge at the band edge and that the maximum attainable SL gain-enhancement is limited by the gain itself. 

In this work, our coupled-Bloch-mode approach is then applied to analyze the threshold condition of various types of PhC lasers. As a first example, we analyze a laser cavity made up of different sections of both passive and active PhC LDWGs. This laser is conceptually similar to the one characterized in \cite{Shinya}. Consistenly with \cite{Shinya}, we show that three different operating regimes can be identified, which can be explained on the basis of the interplay between the distributed feedback in the active region and the passive mirrors. This reveals the great impact of coherent distributed feedback effects in these PhC lasers. As a second example, we analyze a typical PhC laser based on a $\mathrm{LN}$ cavity, such as that of \cite{Xue2016}. In this case, the laser cavity is an active PhC LDWG bounded on either side by classical PhC mirrors. As a last example, we examine a structure simply consisting of an active section, with zero back reflections from the interfaces with the passive output waveguides. Interestingly, in this type of structure the distributed feedback, caused entirely by the gain perturbation in the active region, is enough to allow lasing with reasonable threshold gain.

The paper is organized as follows: in Section II, we present our model and discuss the peculiar characteristics of the self- and cross-coupling coefficients of an active PhC LDWG. In Section III, we present the numerical results applied to both active PhC waveguides and lasers. In Section IV, we finally draw the conclusions.

\section{Numerical model}
\begin{figure}[ht]
	\centering\includegraphics[width=\linewidth]{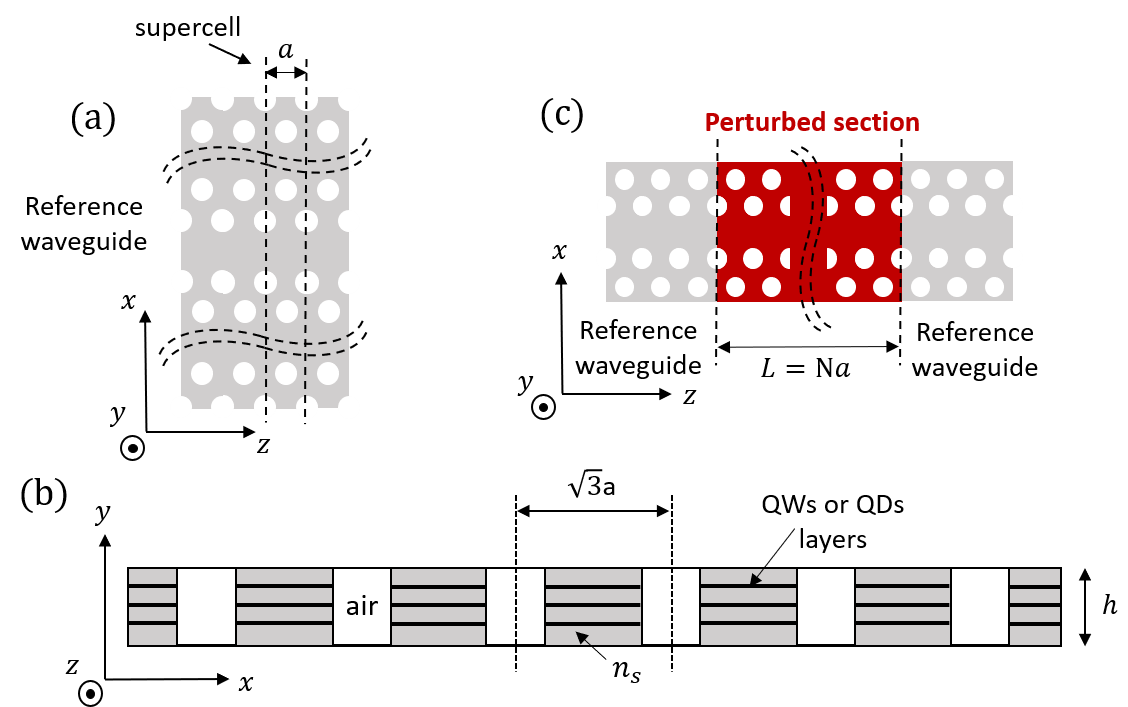}
	\caption{\label{fig:Supercell} Reference PhC waveguide supercell (a). \uline{Cross section of the reference waveguide  supercell at either the input or output plane. The black lines represent QWs or QDs layers} (b). Finite-length, perturbed section in red and reference waveguide in grey (c).}
\end{figure}
\begin{figure}[ht]
	\centering\includegraphics[width=0.75\linewidth]{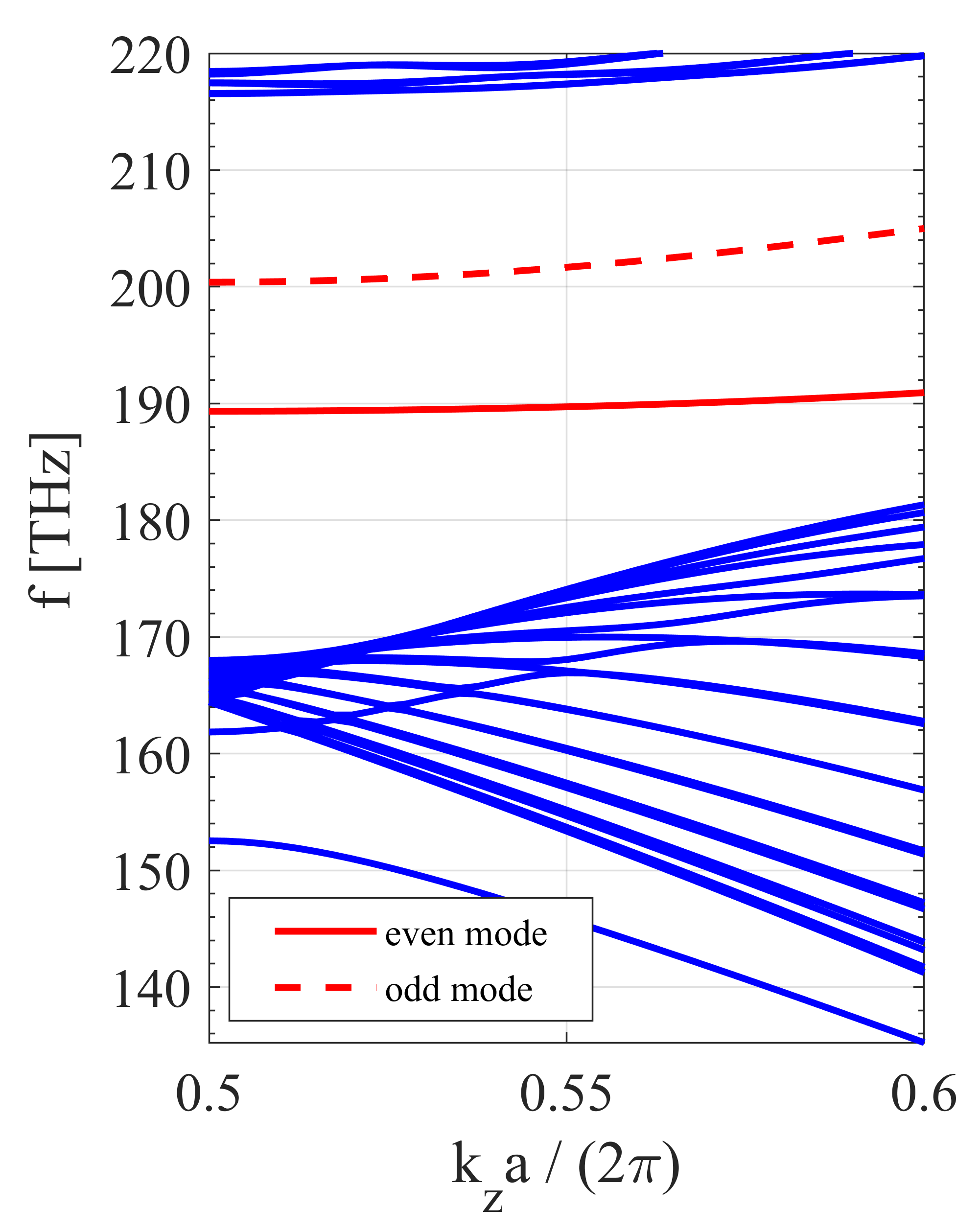}
	\caption{\label{fig:BandStructure_ns3171} Projected band structure of the reference PhC waveguide TE-like modes. The membrane refractive index is $n_s = 3.171$, the lattice constant $a = \mathrm{438 \,nm}$ \uline{and the thickness $h = \mathrm{250 \,nm}$; the other parameters are those of the PhC LDWG on which the lasers in \cite{Xue2016} are based. The membrane is assumed to be suspended in air.} The fundamental guided mode is even with respect to $x$ (red, solid line), the other mode is odd (red, dashed line). \uline{The blue lines represent a subset of the continuum of photonic crystal slab modes found by MPB.}}
\end{figure}
We consider an active, PhC LDWG of finite length bounded on either side by semi-infinite, passive PhC waveguides, as illustrated in Fig.~\ref{fig:Supercell}(c). The passive sections have the same lattice constant and membrane thickness as the active section. \uline{In this context, for \textit{membrane} we simply mean a thin layer of semiconductor, suspended in a low refractive index medium and periodically patterned with holes \cite{DeRossi2009}, as shown in cross-section in Fig.~\ref{fig:Supercell}(b)}. \uline{The \textit{material}} gain \uline{$g$} in the active section may be provided by layers of quantum wells (QWs) or QDs embedded within the PhC membrane, \uline{shown as black lines in Fig.~\ref{fig:Supercell}(b)}. For simplicity, the entire membrane is assumed to contain active material, as in the case of optically pumped waveguides \cite{ek2014a} and lasers based on $\mathrm{LN}$ cavities \cite{Xue2015,Xue2016}, which lack a structure for lateral carrier confinement. However, we note that the coupled-Bloch-mode approach is also applicable to structures with a buried active region, such as LEAP lasers. We consider the gain of the active section and the variation of its refractive index with respect to that of the passive sections as a weak perturbation.
\begin{figure*}[ht]
	\begin{subfigure}[b]{0.33\linewidth}
		\centering\includegraphics[width=\linewidth]{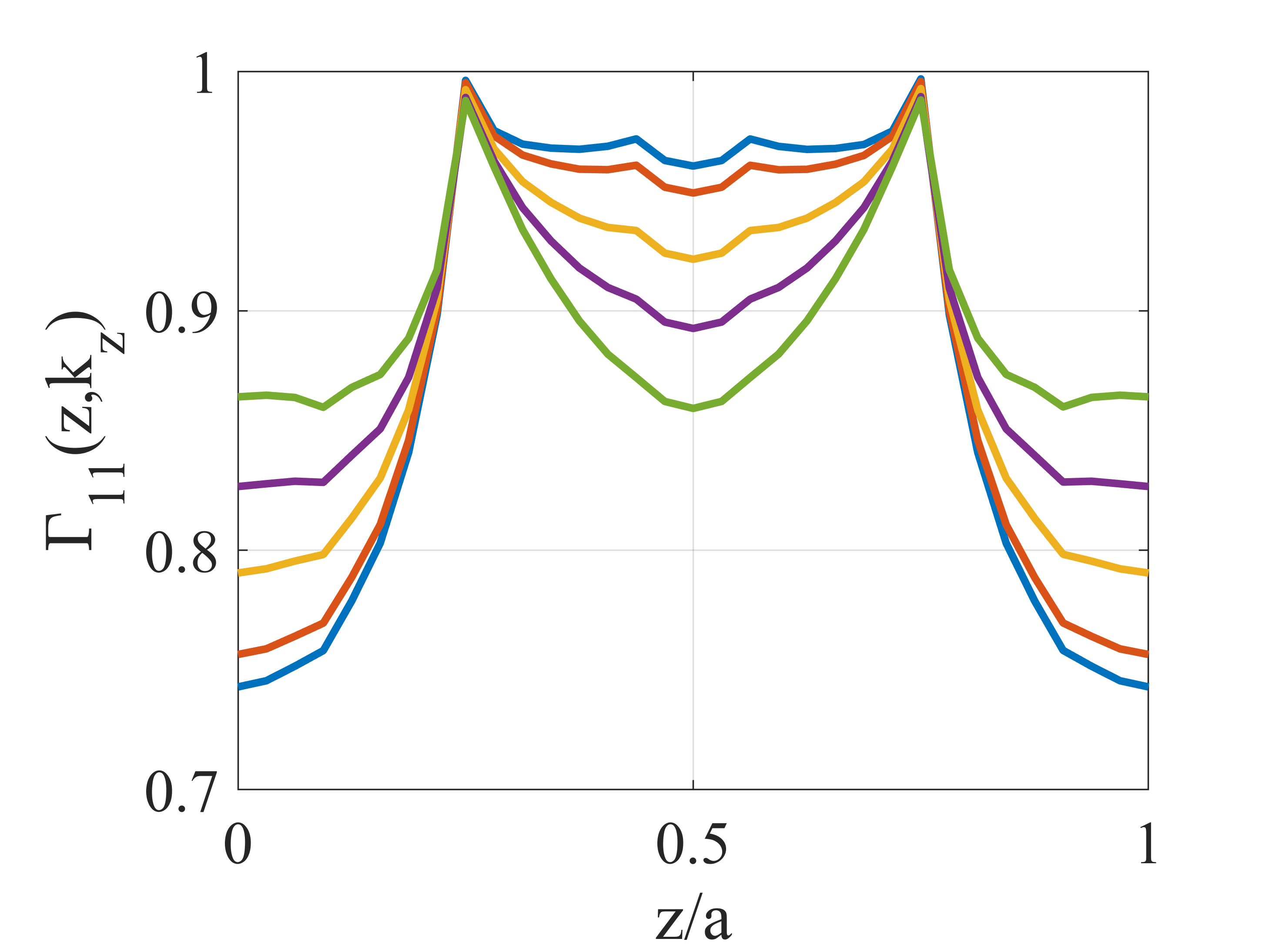}
		\caption{\label{fig:ConfFacta}}
	\end{subfigure}%
	\begin{subfigure}[b]{0.33\linewidth}
		\centering\includegraphics[width=\linewidth]{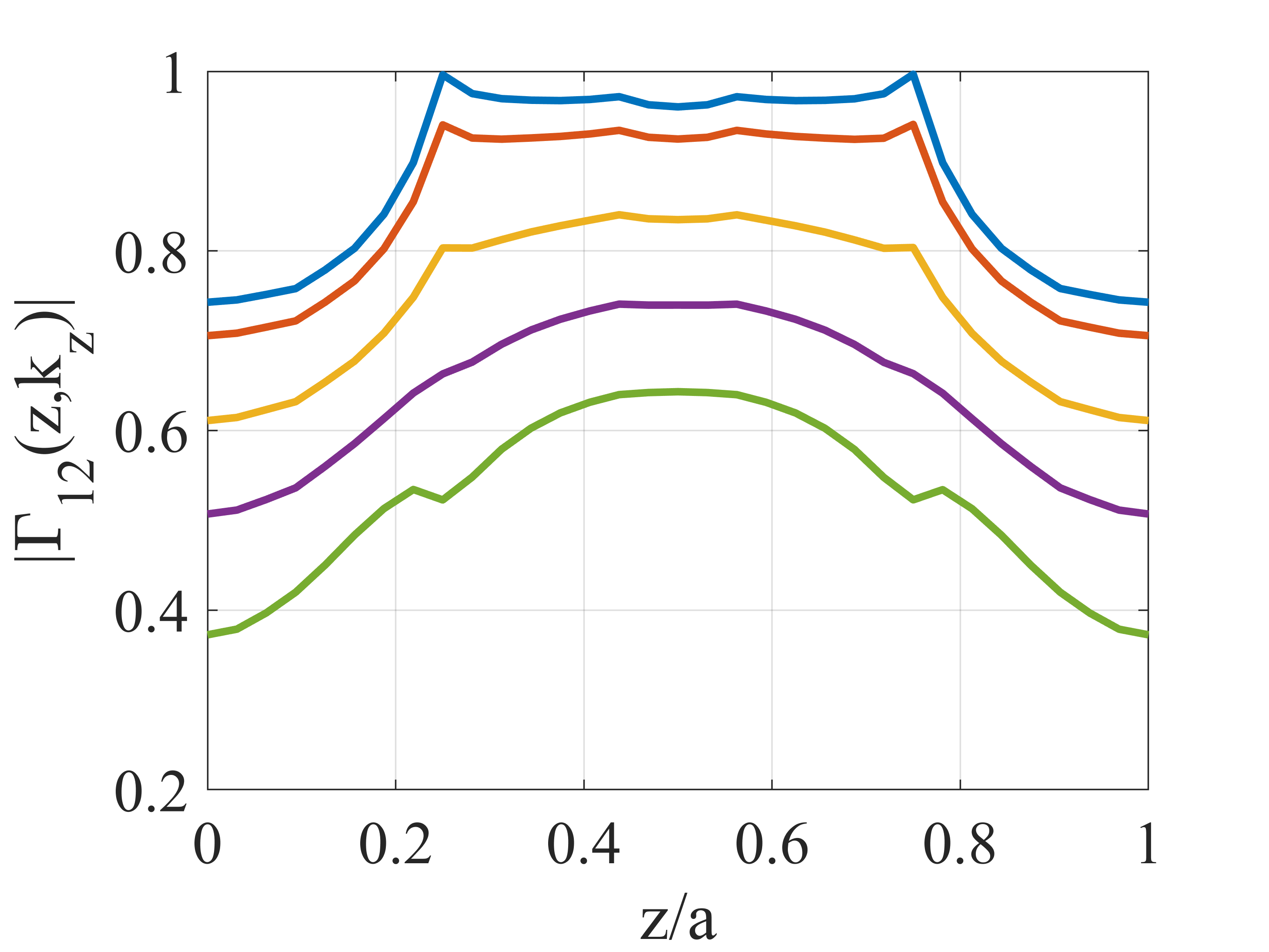}
		\caption{\label{fig:ConfFactb}}
	\end{subfigure}
	\begin{subfigure}[b]{0.33\linewidth}
		\centering\includegraphics[width=\linewidth]{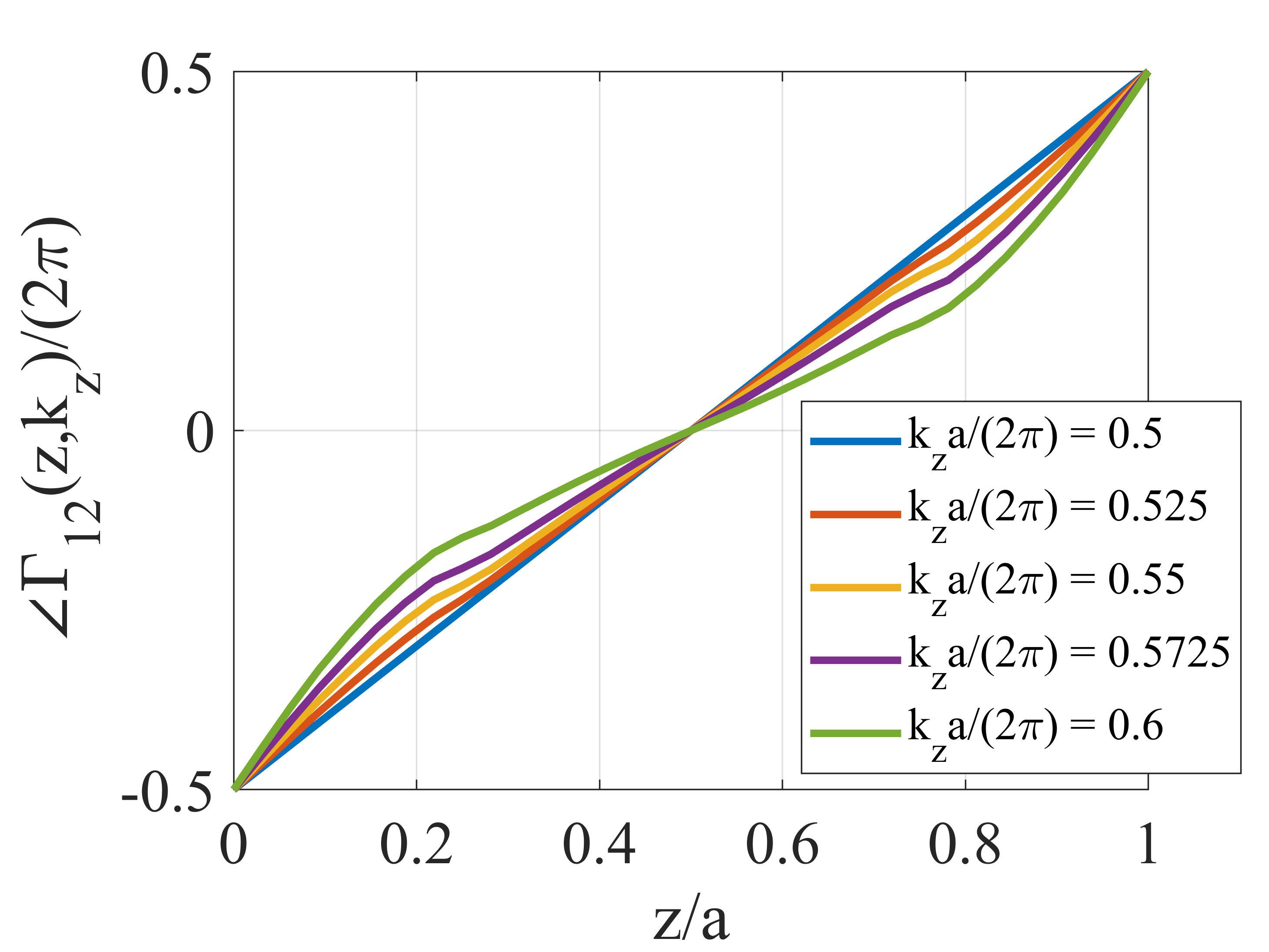}
		\caption{\label{fig:ConfFactc}}
	\end{subfigure}%
	\caption{\label{fig:ConfFact} \sout{confinement factors} \uline{Spatial dependence of} $\Gamma_{11}$ and $\Gamma_{12}$ of the reference waveguide at various $k_z$ values. The membrane refractive index is $n_s = 3.171$; the other parameters are those of the PhC LDWG on which the lasers in \cite{Xue2016} are based. (\ref{fig:ConfFacta}) $\Gamma_{11}$. (\ref{fig:ConfFactb}) Magnitude of $\Gamma_{12}$. (\ref{fig:ConfFactc}) Phase of $\Gamma_{12}$.}
\end{figure*}
The forward- (+) and backward-propagating \mbox{(-)} Bloch modes of the reference waveguide in the frequency-domain are denoted by $\mathbf{E}_{0,\pm}(\mathbf{r},\omega) = \mathbf{e}_{0,\pm}(\mathbf{r},\omega) e^{\pm ik_z(\omega)z}$, where $\mathbf{e}_{0,\pm}(x,y,z,\omega) = \mathbf{e}_{0,\pm}(x,y,z+a,\omega)$, with $k_z$ being the Bloch wave number along the length of the waveguide ($z$) and $a$ the PhC lattice constant. In the active section, the complex refractive index variation compared to the reference waveguide is $\Delta n_s +i\Delta n_i$, where $\Delta n_s$ is the variation of refractive index and $\Delta n_i$ reflects the \uline{\textit{modal}} gain coefficient $g_0$, with $\Delta n_i = -(c/\omega) g_0/2$. \uline{Here, $g_0$ is given by $\Gamma_y g$, where $\Gamma_y$ is the optical confinement factor of the considered electric field guided mode along the y-direction and within the QWs or QDs layers.} This polarization perturbation couples to each other in the active waveguide the otherwise independent Bloch modes of the reference waveguide. Therefore, in the limit of a weak perturbation, the electric field in the active section can be expanded in the basis of the Bloch modes of the reference waveguide, with slowly-varying amplitudes $\psi_{\pm}(z,\omega)$ caused by the coupling. The electric field in the active section thus reads as $\mathbf{E}(\mathbf{r},\omega) = \psi_{+}(z,\omega)\mathbf{E}_{0,+}(\mathbf{r},\omega) + \psi_{-}(z,\omega)\mathbf{E}_{0,-}(\mathbf{r},\omega)$. At a given $\omega$, by neglecting nonlinear effects, two coupled differential equations in $\psi_{\pm}$ are derived \cite{Chen2015}:
\begin{equation}
\label{eq:CoupledBlochModeEquations}
\begin{aligned}
\partial_z\psi_+(z) & = i\kappa_{11}(z)\psi_+(z) + i\kappa_{12}(z) e^{-2ik_z z} \psi_-(z)\\
-\partial_z\psi_-(z) & = i\kappa_{21}(z) e^{+2ik_z z}\psi_+(z) + i\kappa_{11}(z)\psi_-(z)  
\end{aligned}
\end{equation} 
The self- and cross-coupling coefficients are written as $\kappa_{11;12;21}(z,\omega) \simeq [\Delta n_s(\omega)+i\Delta n_i(\omega)](\omega/c)[n_{g,0}(\omega)/n_s]\Gamma_{11;12;21}(z,\omega)$, where $n_{g,0}$ and $n_s$ are the group index and the membrane material refractive index of the reference waveguide. $\Gamma_{11;12;21} (z,\omega)$ are given by
\begin{equation}
\label{eq:ConfFact}
\begin{aligned}
\Gamma_{11}(z,\omega) & = \frac{a\int_{S} \epsilon_0 n_s^2|\textbf{e}_{0}(\textbf{r},\omega)|^2 F(\textbf{r})dS}{\int_{V} \left[\epsilon_0 n_b^2(\textbf{r}) |\textbf{e}_0(\textbf{r},\omega)|^2\right] dV}\\ \Gamma_{12}(z,\omega) & = \frac{a\int_{S} \epsilon_0 n_s^2\left[\textbf{e}_{0,-}(\textbf{r},\omega) \cdot \textbf{e}^*_{0,+}(\textbf{r},\omega)\right] F(\textbf{r})dS}{\int_{V} \left[\epsilon_0 n_b^2(\textbf{r}) |\textbf{e}_0(\textbf{r},\omega)|^2\right] dV}
\end{aligned}
\end{equation}
with $\Gamma_{21} = \Gamma_{12}^*$. Here, $V$ is the volume of the supercell (see Fig.~\ref{fig:Supercell}(a)), $S$ the transverse section at position $z$ and $n_b (\mathbf{r}) = n_s (=n_{\mathrm{holes}})$ in the membrane (holes) is the reference waveguide background refractive index. The spatial distribution of the complex perturbation is taken into account by $F(\mathbf{r})$, which is equal to 1 (0) in the membrane (holes) of the active section. \uline{$\Gamma_{11}$ is the ratio between the electric field energy in the gain region of a supercell and that stored in the whole supercell, by assuming the modal gain $g_0$ to be uniform through the semiconductor slab.} 
\begin{figure}[ht]
	\centering\includegraphics[width=\linewidth]{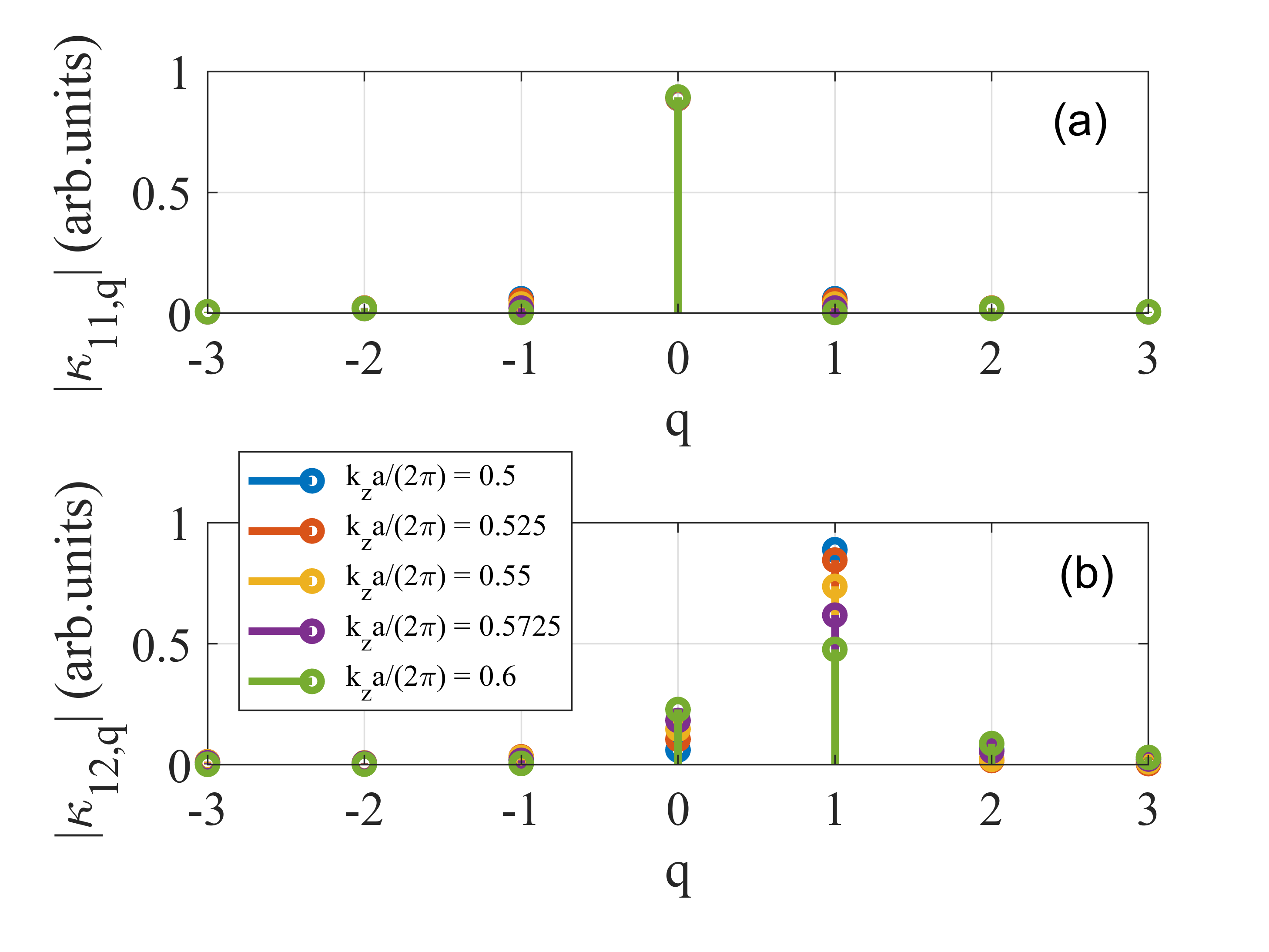}
	\caption{\label{fig:Gamma11_12Harmonics} Magnitude of the Fourier components of the self-coupling $\kappa_{11}(z)$ (a) and cross-coupling coefficient $\kappa_{12}(z)$ (b) at various $k_z$ values.}
\end{figure}
\uline{We note that the self-coupling coefficient is caused by the fact that the modes used as a basis for the electric field see, in the perturbed waveguide, an average complex refractive index profile which is different from that of the reference waveguide where the modes are defined}. \sout{We note that,} Due to the $z-$periodicity of $\mathbf{e}_{0,\pm}$ and $F(\mathbf{r})$, \sout{confinement factors} $\Gamma_{11}$ and $\Gamma_{12}$ and, therefore, the coupling coefficients are periodic with $z$. As an example, we consider a reference waveguide with refractive index $n_s = 3.171$; the other parameters are those of the PhC waveguide on which the lasers in \cite{Xue2016} are based, with $a = 438\mathrm{\,nm}$, \uline{$n_{\mathrm{holes}} = 1$ and membrane thickness $h = 250\mathrm{\,nm}$. The membrane is assumed to be suspended in air, consistently, for instance, with \cite{Xue2015,Xue2016} or \cite{MatsuoNatPhot2013,MatsuoJSTQE2013}, which investigate air-bridge structures}. Band structure and TE-like Bloch modes of the reference waveguide are computed by the plane wave eigensolver MIT Photonic-Bands (MPB) \cite{MPBpaper} (Fig.~\ref{fig:BandStructure_ns3171}). \uline{The blue lines represent a subset of the continuum of photonic crystal slab modes found by MPB \cite{skorobogatiy_yang_2008}. These modes are confined to the slab along the y-direction, but are delocalized in the x- and z-direction}. We apply the coupled-Bloch-mode approach to the fundamental guided mode (red, solid line), \uline{which is confined both along the x- and y-direction}. The magnitude and phase of $\Gamma_{11}$ and $\Gamma_{12}$ over a unit cell are displayed in Fig.~\ref{fig:ConfFact} at various $k_z$ values. Given the spatial distribution of the perturbation (that is $F(\textbf{r})$), they only depend on the reference waveguide geometry and Bloch modes and not on the magnitude of the perturbation. For uniformly pumped membranes, \sout{confinement factors} $\Gamma_{11}$ and $\Gamma_{12}$ are intrinsic parameters of the reference waveguide and they determine how strong the cross-coupling is with respect to self-coupling. The coupling coefficients are indeed proportional to \sout{confinement factors} $\Gamma_{11}$ and $\Gamma_{12}$ through $n_{g,0}$ and the complex refractive index perturbation \sout{and they are periodic in $z$ with the same periodicity of $\Gamma_{11,12,21}$}. Therefore, they can be expanded in a Fourier series as $\kappa_{11;12}(z,\omega) = \sum_q \kappa_{11,q;12,q}(\omega) \exp(+iq2\pi z/a)$. Since the phase of $\Gamma_{12}$ is approximately linear with $z$ with a slope equal to $2\pi/a$, $\kappa_{12,q=1}$ is proportional to $\left\langle |\kappa_{12}(z)|\right\rangle$, which is comparable with $\kappa_{11,q=0}$. This is illustrated in Fig.~\ref{fig:Gamma11_12Harmonics}, showing the Fourier components of $\kappa_{11}(z)$ and $\kappa_{12}(z)$  in arbitrary units at various $k_z$ values. Interestingly, we find that the complex refractive index perturbation not only produces a self-coupling coefficient \sout{(as expected in any standard waveguide)} \uline{proportional to the average of the perturbation (i.e. the average of $\Gamma_{11}(z)$ in Fig.~\ref{fig:ConfFact}(a))}, but also a strong cross-coupling, whose magnitude is close to that of the self-coupling. This strong cross-coupling is possible thanks to the linear phase of $\Gamma_{12}(z)$; if this linear phase component were not present, the cross-coupling would be negligible. By analyzing the contribution of the various field components ($\hat{x}$, $\hat{y}$ and $\hat{z}$ component) to $\Gamma_{12}(z)$, we have found that the linear phase variation is caused by a significant $\hat{z}$ component of the TE-like guided mode, which is typically negligible in standard waveguides. Here the non-negligible $\hat{z}$ component is due to the strong lateral (\sout{$y-$} \uline{$x-$}direction) confinement obtained by the PhC. 

Having now more insight into the main properties of the coupling coefficients, we derive an analytical expression for the transmission matrix describing the field propagation over a unit cell. As illustrated in Fig.~\ref{fig:Gamma11_12Harmonics}, all harmonics in the self-coupling coefficient, other than $q=0$, are negligible; similarly, the harmonic with $q = 1 (=-1)$ is the dominant one in the cross-coupling coefficient $\kappa_{12}$ ($\kappa_{21}$). This is true over a wide frequency range and especially close to the band edge. Therefore, by only retaining the dominant harmonics, Eqs.~(\ref{eq:CoupledBlochModeEquations}) are turned into 
\begin{equation}
\label{eq:CoupledBlochModeEquationsFourier}
\begin{aligned}
\partial_z\psi_+ & \simeq i\kappa_{11,q=0}(\omega)\psi_+ + i\kappa_{12,q=1}(\omega) e^{+2i\delta(\omega) z} \psi_-\\
-\partial_z\psi_- & \simeq i\kappa_{21,q=-1}(\omega) e^{-2i\delta(\omega) z}\psi_+ + i\kappa_{11,q=0}(\omega)\psi_-  
\end{aligned}
\end{equation}     
where $\delta(\omega) = \pi/a - k_z(\omega)$ is the detuning from the band edge. \uline{By defining $b^{\pm}(z,\omega) = \psi_{\pm}(z,\omega)\exp\{\mp i\delta(\omega)z\}$, the system of Eqs.~(\ref{eq:CoupledBlochModeEquationsFourier}) is turned into a pair of partial differential equations with z-independent coefficients. This allows to analytically solve it over a unit cell as an initial value problem. That is, $b^{\pm}(z_0 + a,\omega)$ are computed by assuming $b^{\pm}(z_0,\omega)$ to be known, with the input coordinate $z_0$ of the unit cell conveniently chosen to be zero. Therefore, the unit cell transmission matrix in terms of $b^{\pm}$ is obtained. However, we need the transmission matrix for $c^{\pm}(z,\omega)$, with $c^{\pm}(z,\omega) = \psi_{\pm}(z,\omega)\exp\{\pm ik_z(\omega)z\}$. Since $b^{\pm}(z,\omega) = c^{\pm}(z,\omega)\exp\{\mp i(\pi/a)z\}$, the unit cell transmission matrix in terms of $c^{\pm}$ is obtained through the change of variables $b^{\pm}(z_0 + a,\omega) = c^{\pm}(z_0 + a,\omega)\exp\{\mp i\pi\}$ and $b^{\pm}(z_0,\omega) = c^{\pm}(z_0,\omega)$. Therefore,} \sout{By putting $\psi_{\pm} = c^\pm e^{\mp ik_z z}$ and solving the resulting equations over a unit cell}, we can relate $c^\pm$ at the input ($N-1$) and output ($N$) of the generic $N_{\mathrm{th}}$ cell by
\begin{equation}
\label{eq:cpm_Ta}
\left[
\begin{aligned}
& c_N^+ (\omega) \\
& c_N^- (\omega)
\end{aligned}
\right]
=
\mathbf{T}_a (\omega)
\left[
\begin{aligned}
& c_{N-1}^+ (\omega) \\
& c_{N-1}^- (\omega)
\end{aligned}
\right] 
\end{equation}
\sout{$[c_N^+ (\omega)   c_N^- (\omega)]^T = \mathbf{T}_a (\omega) [c_{N-1}^+ (\omega)   c_{N-1}^- (\omega)]^T$,} where $\mathbf{T}_a$ is the unit cell transmission matrix \uline{in terms of $c^{\pm}$} \sout{and $T$ denotes the transpose operator}. The elements of $\mathbf{T}_a$ are given by 
\begin{equation}
\label{eq:Tmatrix}
\begin{aligned}
T_{a,11;a,22} & = -\cosh\left(\gamma a\right) \mp i\left(\tilde{\delta}/\gamma\right)\sinh\left(\gamma a\right) \\ 
T_{a,12;a,21} & = \mp i\left(\kappa_{12,q=1;21,q=-1}/\gamma\right)\sinh\left(\gamma a\right) 
\end{aligned}
\end{equation}
with $\tilde{\delta}(\omega) = \kappa_{11,q=0}(\omega) - \delta(\omega)$ and $\gamma(\omega) = \sqrt{\kappa_{12,q=1}(\omega)\kappa_{21,q=-1}(\omega) - \tilde{\delta}^2(\omega)}$. This approach allows to directly relate the eigenvalues of $\mathbf{T}_a$ to the coupling coefficients and the detuning. In fact, the eigenvalues $\lambda_{1,2}$ are readily obtained as $\lambda_{1,2} = \exp\{\pm\mathrm{Re}\left\lbrace\gamma a\right\rbrace\} \exp\{\pm i[\mathrm{Im}\left\lbrace\gamma a\right\rbrace +\pi]\}$. From here, we define $\pm g_{\mathrm{eff}} = \pm2\mathrm{Re}\left\lbrace\gamma\right\rbrace$ as the effective gain per unit length, representing the gain experienced by the forward ($+$) and backward ($-$) Bloch modes of the \textit{active} section while propagating in a unit cell. Similarly, $\pm \beta_{\mathrm{eff}}a = \pm[\mathrm{Im}\left\lbrace\gamma a\right\rbrace +\pi]$ is the phase shift per cell of these Bloch modes. From this phase shift, we can then calculate the group index of the Bloch modes of the active waveguide as $n_{g}(\omega) = \mathrm{c}\:d\beta_{\mathrm{eff}}/d\omega$. By applying Frobenius theorem \cite{Frobenius}, the transmission matrix of an active waveguide of $N$ unit cells is computed as 
\begin{equation}
\label{eq:FrobeniusTheorem}
\mathbf{T}^N_a = \mathbf{M}\mathbf{\lambda_a}^N\mathbf{M}^{-1} 
\end{equation}
where $\mathbf{M}$ and $\mathbf{\lambda_a}$ are given by
\begin{equation}
\label{eq:M_Lambda}
\mathbf{M} = 
\left[
\begin{aligned}
u_{11} & \quad u_{12} \\
u_{21} & \quad u_{22}
\end{aligned}
\right], 
\quad
\mathbf{\lambda_a} = 
\left[
\begin{aligned}
\lambda_1 & \quad 0 \\
0 & \quad \lambda_2
\end{aligned}
\right] 
\end{equation}
with $\mathbf{u}_1 = \left[u_{11} \quad u_{21}\right]^{T}$ and $\mathbf{u}_2 = \left[u_{12} \quad u_{22}\right]^{T}$ being the eigenvectors of $\mathbf{T}_a$ \uline{and $T$ denoting the transpose operator}. As well as being numerically efficient, this approach for computing $\mathbf{T}^N_a$ allows for a useful physical interpretation: $\mathbf{\lambda_a}^N$ can be seen as the transmission matrix describing the propagation of the Bloch modes of the \textit{active} section over the $N$ unit cells. $\mathbf{M}^{-1}$ and $\mathbf{M}$ can be interpreted as the transmission matrices of the interfaces between the active section and, respectively, the left and right passive waveguides; they physically account for the mismatch between the Bloch modes of the active and passive sections. \uline{By applying the relationships between a transmission and a scattering matrix \cite{Coldren}, the transmission matrix $\mathbf{T}^N_a$ can be turned into the corresponding scattering matrix. The scattering parameters are given by}
\begin{equation}
\label{eq:Smatrix}
\begin{aligned}
S_{11} & = \frac{\left(\lambda_1^N - \lambda_2^N\right)T_{a,21}}{\left(\lambda_1^N - \lambda_2^N\right)T_{a,11} + \left(\lambda_2^{N+1} - \lambda_1^{N+1}\right)} \\ 
S_{12;21} & = \frac{\left(\lambda_2 - \lambda_1\right)}{\left(\lambda_1^N - \lambda_2^N\right)T_{a,11} + \left(\lambda_2^{N+1} - \lambda_1^{N+1}\right)} \\
S_{22} & = \frac{\left(\lambda_2^N - \lambda_1^N\right)T_{a,12}}{\left(\lambda_1^N - \lambda_2^N\right)T_{a,11} + \left(\lambda_2^{N+1} - \lambda_1^{N+1}\right)}
\end{aligned}
\end{equation} 

As outlined in \cite{Chen2015}, the coupled-Bloch-mode approach breaks down at large values of $g_0$. Specifically, the larger $n_{g,0}$ becomes, the smaller is the value of the gain coefficient at which the model starts to fail. As a further limitation, here we also point out that the model is actually not applicable at the band edge of the reference waveguide, independently of the value of $g_0$. This is because the group velocity of the Bloch modes used as a basis for the field expansion is ideally equal to zero at the band edge of the reference waveguide. These considerations imply that, in principle, we could not analyze active PhC waveguides in the frequency range close to the critical point $k_z = \pi/a$. However, this limitation can be overcome with the following trick: to analyze an active PhC waveguide including the frequency range of the band edge and some frequencies in the stop-band, we start from a reference waveguide with index $n_s$ slightly \textit{larger} than that of the waveguide to be studied. The waveguide of interest is then seen as having a real refractive index perturbation $\Delta n_s<0$ with respect to the reference one; therefore, its dispersion relation is shifted to higher frequencies with respect to the band edge of reference waveguide, where the model is not applicable. This approach will be \sout{illustrated} \uline{applied} in section IIIA.

\section{Simulation results}

\subsection{Line-defect active waveguide}

\begin{figure}
	\centering\includegraphics[width=\linewidth]{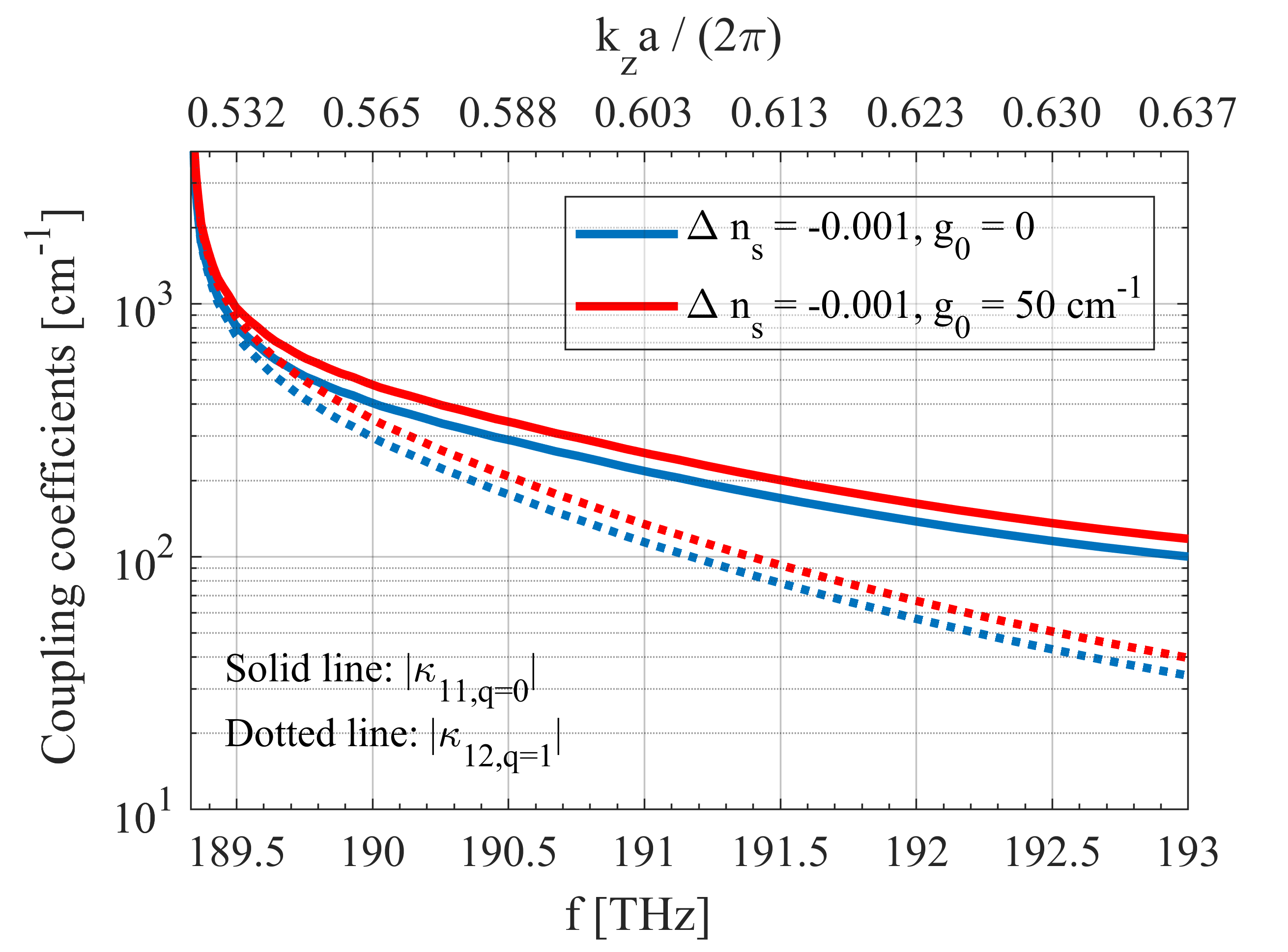}
	\caption{\label{fig:CouplCoeff}Magnitude of self- ($\kappa_{11,q=0}$) and cross-coupling coefficient ($\kappa_{12,q=1}$) for $g_0 = 0$ (blue) and $g_0 = 50 \,\mathrm{cm^{-1}}$ (red). In both cases, $\Delta n_s = -0.001$.}
\end{figure}
Fig.~\ref{fig:CouplCoeff} shows the magnitude of the self- ($\kappa_{11,q=0}$) and cross- coupling coefficient ($\kappa_{12,q=1}$) for $g_0 = 0$ (blue) and $g_0 = 50 \,\mathrm{cm^{-1}}$ (red). In both cases, $\Delta n_s = -0.001$. This figure proves that the cross-coupling coefficient is always comparable to the self-coupling coefficient. This is consistent with results also shown in \cite{MichaelisPR}. 
However, as compared to \cite{MichaelisPR}, we have clarified here the physical origin of this peculiar behaviour. We also observe that the coupling coefficients depend on the intensity of the perturbation, as expected, but also on frequency and they significantly increase as the frequency approaches the band edge as consequence of the SL effect. Therefore, Bloch modes at smaller frequency and/or with higher gain of the active section experience stronger distributed feedback as compared to higher frequency Bloch modes and/or lower active section gain.  

\begin{figure}
	\centering\includegraphics[width=\linewidth]{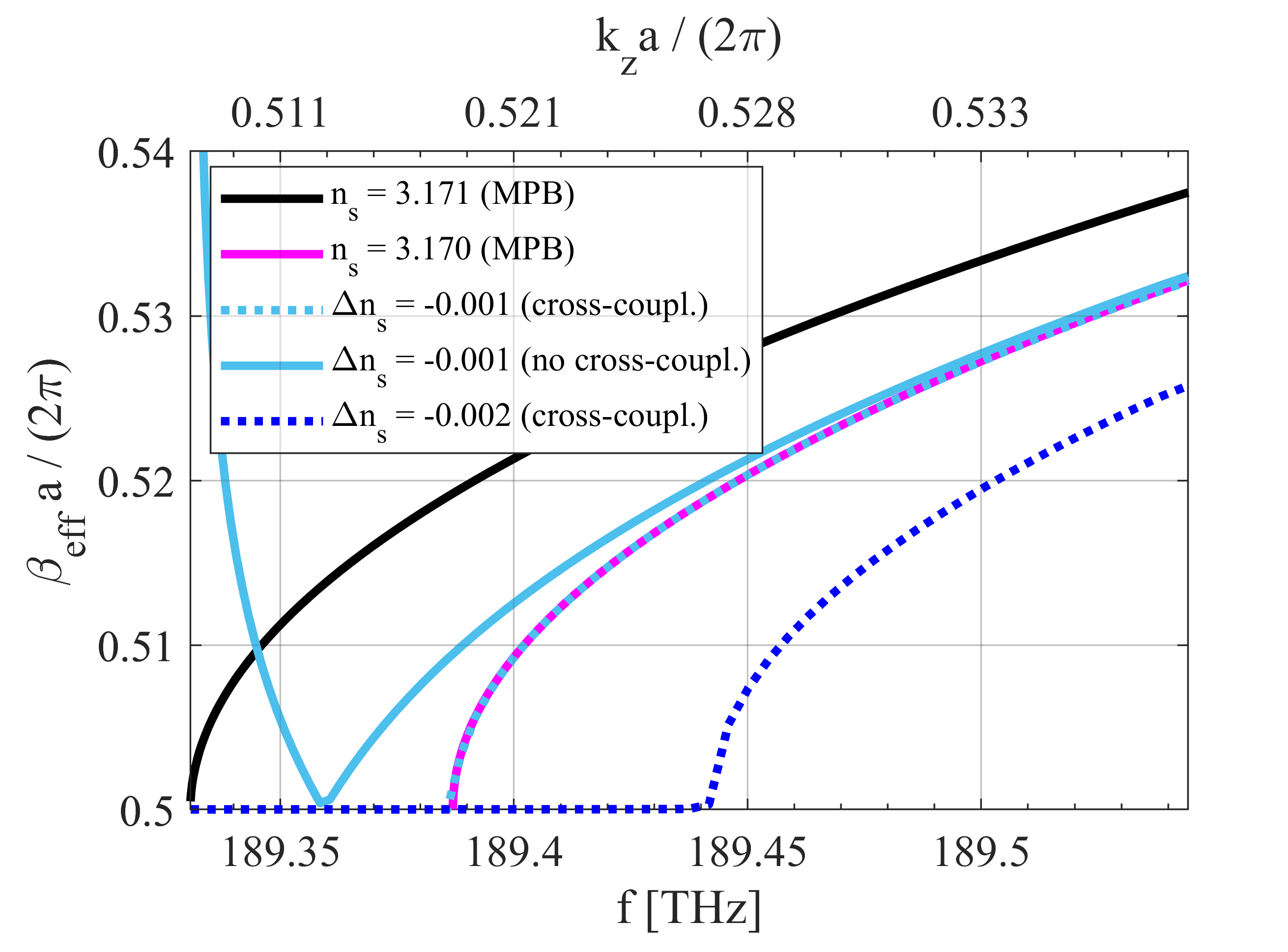}
	\caption{\label{fig:gammaIM_MPBcomp} Dispersion curve computed by MPB for $n_s = 3.171$ (black) and $n_s = 3.170$ (pink). Phase shift per cell of the Bloch modes of the perturbed waveguide with (light-blue, dotted) and without (light-blue, solid) cross-coupling; the reference waveguide has $n_s = 3.171$ and the real refractive index perturbation is $\Delta n_s = -0.001$, with $g_0 = 0$. The dark-blue, dotted curve is for $\Delta n_s = -0.002$, $g_0 = 0$.} 
\end{figure}
\begin{figure}
	\centering\includegraphics[width=\linewidth]{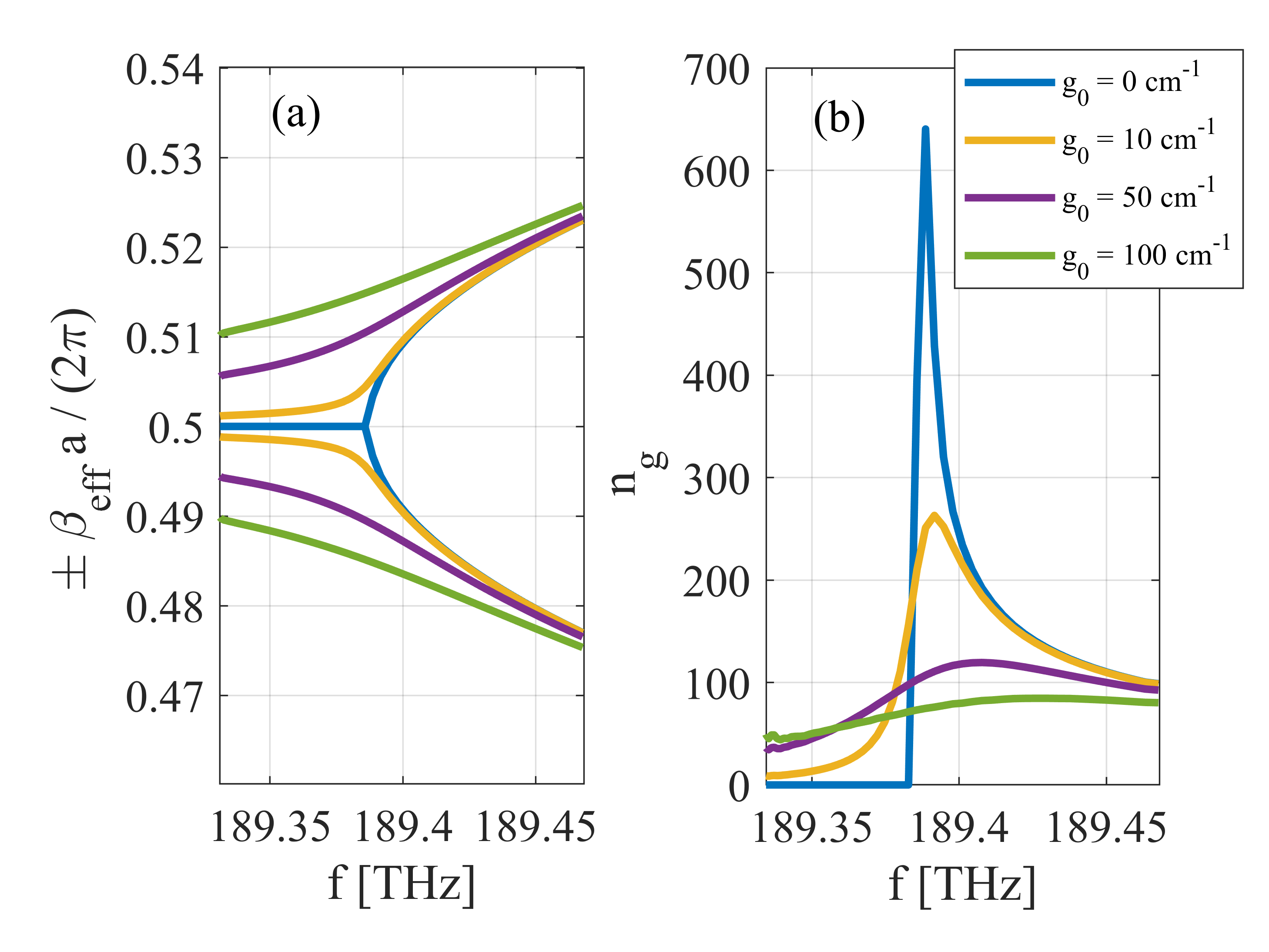}
	\caption{\label{fig:gammaIM_ngPert} Phase shift per cell of the Bloch modes of the active waveguide (a) and corresponding group index (b) at various gain values. The reference waveguide has $n_s = 3.171$ and the real refractive index perturbation is $\Delta n_s = -0.001$.}
\end{figure}
In this section, we analyze how a generally complex refractive index perturbation impacts the dispersion relation of a PhC waveguide. To validate our model, we first study the case of a purely real refractive index perturbation, because the results obtained by our approach can be compared with MPB simulations. We consider a reference passive waveguide with  $n_s = 3.171$ and we report in black in Fig.~\ref{fig:gammaIM_MPBcomp} the corresponding dispersion relation computed by MPB. We then consider a small, real refractive index perturbation $\Delta n_s<0$ and we compare the dispersion relation calculated with our model (i.e. $\beta_{\mathrm{eff}}(\omega)a$) with that calculated by MPB for a passive waveguide with refractive index $n_s+\Delta n_s$. As $n_s$ decreases, the dispersion curve shifts to higher frequencies, meaning that the Bloch modes become evanescent at lower frequencies. The formation of this stopband for \textit{Bloch} modes is correctly reproduced by our approach: the light-blue, dotted curve  in Fig.~\ref{fig:gammaIM_MPBcomp} is the dispersion relation calculated by our approach for $\Delta n_s = -0.001$ and it perfectly overlaps with that obtained by MBP for $n_s=3.170$. If the contribution of the cross-coupling coefficients in Eqs.~(\ref{eq:CoupledBlochModeEquationsFourier}) is neglected, we find the dispersion relation shown by the light-blue, solid curve in Fig. \ref{fig:gammaIM_MPBcomp}; in this case, the Bloch modes of the perturbed waveguide are not evanescent in the stop-band and the dispersion relation disagrees with the MPB prediction. These results validate the correctness of our approach and emphasize the role of the cross-coupling terms. Furthermore, the possibility to analyze the propagation of the Bloch-modes in a portion of the stop-band of the perturbed, passive waveguide will be exploited in the next section to study laser configurations similar to the LEAP lasers of \cite{Matsuo2012,Matsuo2010}. 

\begin{figure}
	\centering\includegraphics[width=\linewidth]{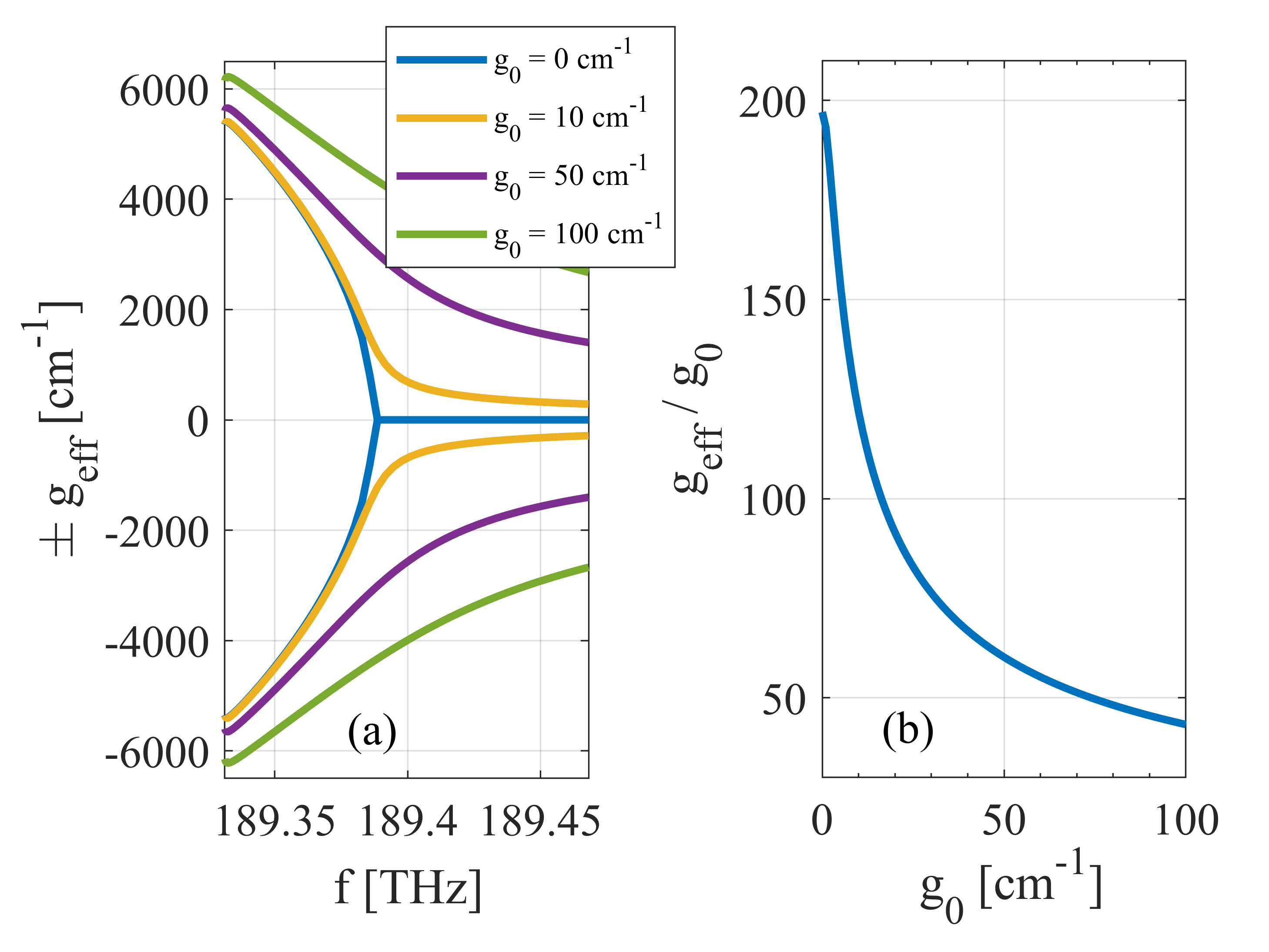}
	\caption{\label{fig:gammaRE_EnhcFact} Effective gain as a function of frequency at various gain values (a). Gain enhancement factor at $f\simeq\mathrm{189.388\,THz}$ (b). The reference waveguide has $n_s = 3.171$ and the real refractive index perturbation is $\Delta n_s = -0.001$.}
\end{figure}    
As a second example, we consider the impact of a gain  perturbation, assuming now $\Delta n_s = -0.001$ and $g_0>0$. Fig.~\ref{fig:gammaIM_ngPert}(a) shows how the dispersion relation  of the Bloch modes of the active waveguide is modified by the gain perturbation and Fig.~\ref{fig:gammaIM_ngPert}(b) reports the corresponding group index at various gain values. For $g_0 = 0$, the group index diverges at $f\simeq\mathrm{189.388\,THz}$, because this frequency corresponds to the band edge of the perturbed waveguide. At lower frequencies, the Bloch modes are evanescent and the group index is zero,  \uline{because evanescent waves do not carry any active power}. As $g_0$ increases, the group index is gradually reduced in the passband, thus limiting the SL effect; on the contrary, it gradually increases in the stopband, where $\beta_{\mathrm{eff}}$ is now different from zero. This is consistent with results reported in \cite{Gric2012}, which were obtained through a non-perturbative approach. The corresponding $g_{\mathrm{eff}}$ versus frequency is shown in Fig.~\ref{fig:gammaRE_EnhcFact}(a) at various  values of $g_0$; we also report in Fig.~\ref{fig:gammaRE_EnhcFact}(b) the gain  enhancement factor $g_{\mathrm{eff}}/g_0$ at $f\simeq\mathrm{189.388\,THz}$ obtained by our approach. Since cross-coupling is not negligible, the gain-induced distributed feedback becomes more and more important as the gain increases, thus causing the decrease of the group index; as a consequence, the gain enhancement factor, which is based on the SL effect, is reduced by increasing the pumping of the active region. This result is consistent with \cite{Gric2012} and confirms that a fundamental limitation to the achievable SL enhancement of the gain is imposed by the gain itself. 

\subsection {PhC Lasers}

In this section, we  apply the coupled-Bloch-mode approach to model the threshold characteristics of various types of PhC laser cavities. The threshold condition is found by calculating the complex loop-gain (LG) of the cavity.
\begin{figure}
	\centering\includegraphics[width=\linewidth]{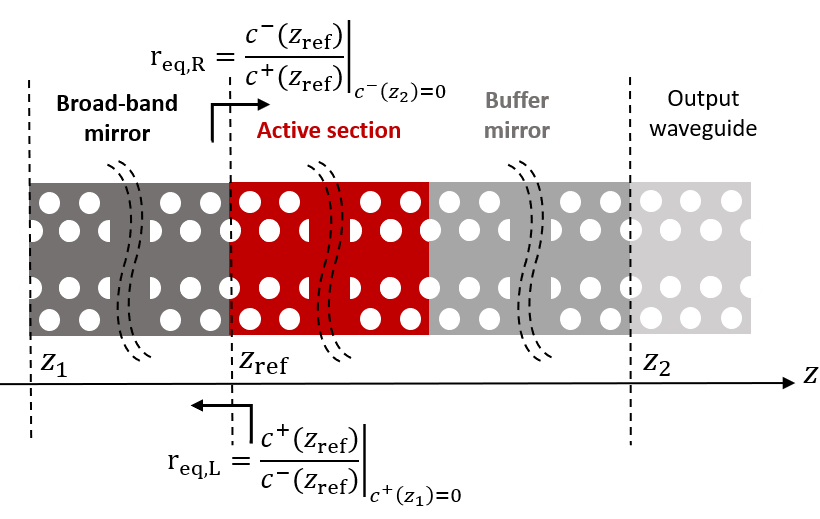}
	\caption{\label{fig:TypeABCcavScheme} Laser cavity made up of three sections, perturbed in refractive index (passive mirrors) and gain (active section) with respect to a reference, passive waveguide with $n_s = 3.171$. The  section to the left of the active region (broad-band mirror) is passive, with $\Delta n_s = -0.002$ and length fixed to $L = 30a$; the active section has $\Delta n_s = -0.001$, $g_0>0$ and $L = 10a$. The section to the right of the active region is also passive and acts as a buffer mirror, with $\Delta n_s = -0.001$ (Type A), $\Delta n_s = -0.0005$ (Type B) or $\Delta n_s = -0.002$ (Type C) and variable length $L_{\mathrm{buffer}}$. The output waveguide coincides with the reference waveguide. \uline{The position of the reference plane to compute the complex loop-gain is denoted by $z_{\mathrm{ref}}$}.}
\end{figure} 

As a first example, we analyze the cavity shown in Fig.~\ref{fig:TypeABCcavScheme}. The cavity is made up of three sections, perturbed in refractive index (passive mirrors) and gain (active section) with respect to a reference, passive waveguide with $n_s = 3.171$. Referring to Fig.~\ref{fig:TypeABCcavScheme}, the  section to the left of the active region (rear broad-band mirror) is passive, with $\Delta n_s = -0.002$ and length fixed to $L = 30a$; the active section has $\Delta n_s = -0.001$, $g_0>0$ and $L = 10a$. The section to the right of the active region (front mirror) is also passive and acts as a buffer mirror, coupling the active section with the reference, output waveguide. Depending on the refractive index perturbation of this buffer region, the band edge of its dispersion relation shifts with respect to that of the active region; on the basis of the buffer refractive index perturbation, we have identified three different types of buffer, thus providing different back reflections to the active region. These will be denoted as Type A (moderate back reflection), Type B (low back reflection) and Type C (high back reflection). These laser cavities are conceptually similar to those in \cite{Shinya}, where the relative shift of the dispersion relation of the passive sections with respect to the active section was obtained by changing the width of the corresponding waveguides \cite{Notomi2001}. For this laser cavity, the LG is computed as the product between the left ($r_{\mathrm{eq,L}}$) and right ($r_{\mathrm{eq,R}}$) field reflectivity at the interface between the active section and the rear mirror (see Fig.~\ref{fig:TypeABCcavScheme}). \uline{Specifically, $r_{\mathrm{eq,L}}$ is readily obtained by Eqs.~(\ref{eq:Smatrix}) as the $S_{22}$ parameter of the broad-band mirror; $r_{\mathrm{eq,R}}$ corresponds to the field reflectivity resulting from the cascade of the active section and buffer mirror scattering matrices, i.e.} 
\begin{equation}
\begin{aligned}
\label{eq:reqR}
r_{\mathrm{eq,R}} & = S_{\mathrm{11,active}} + r_{\mathrm{eq,2}} \\
r_{\mathrm{eq,2}} & = \frac{S_{\mathrm{12,active}}S_{\mathrm{11,buffer}}S_{\mathrm{21,active}}}{1-S_{\mathrm{11,buffer}}S_{\mathrm{22,active}}}
\end{aligned}
\end{equation}
with $S_{ij,\mathrm{buffer}}$ and $S_{ij,\mathrm{active}}$ denoting the buffer and active section scattering parameters computed by Eqs.~(\ref{eq:Smatrix}). The longitudinal resonant modes are those satisfying  $\angle \mathrm{LG}=2m\pi$; the threshold gain $g_{\mathrm{0,th}}$ is the smallest $g_0$ value ensuring  $|\mathrm{LG}|=1$ at the frequency of the longitudinal modes. \sout{$r_{\mathrm{eq,R,L}}$ are calculated from the transmission matrices of the active section and buffer mirror as computed by Eqs.~(\ref{eq:FrobeniusTheorem})}. 
\begin{figure}
	\centering\includegraphics[width=\linewidth]{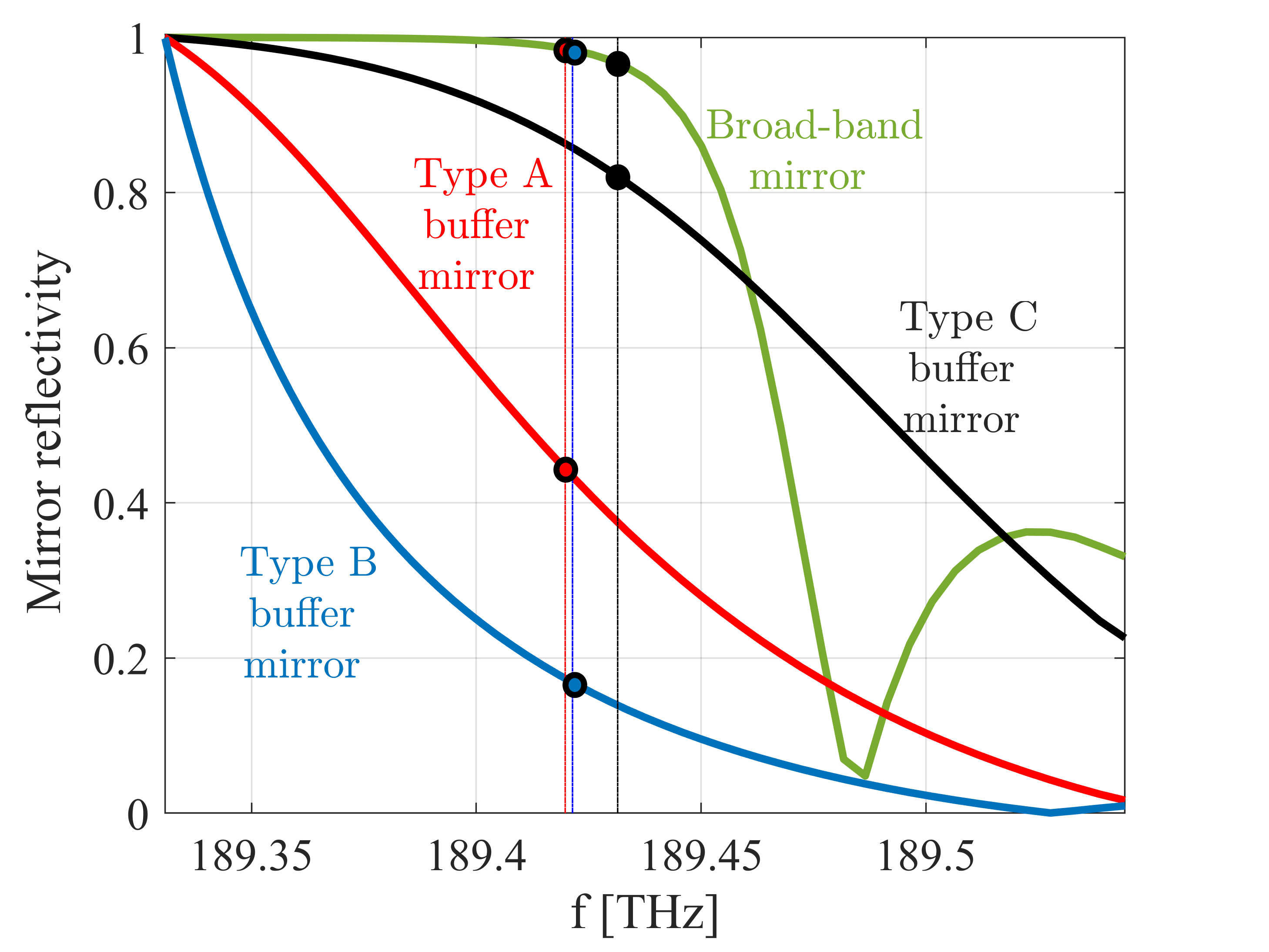}
	\caption{\label{fig:MirrorsSpectra} Broad-band mirror reflectivity; $\Delta n_s = -0.002$ and $L=30a$ (green). Buffer mirror reflectivity for $L_{\mathrm{buffer}}=15a$ and $\Delta n_s = -0.002$ (type C, black), $\Delta n_s = -0.001$ (type A, red) and $\Delta n_s = -0.0005$ (type B, blue). The bullets indicate the position of the lasing mode, in the three different operating regimes, both on the broad-band and buffer mirror reflection spectrum.}
\end{figure}    
\begin{figure}
	\centering\includegraphics[width=\linewidth]{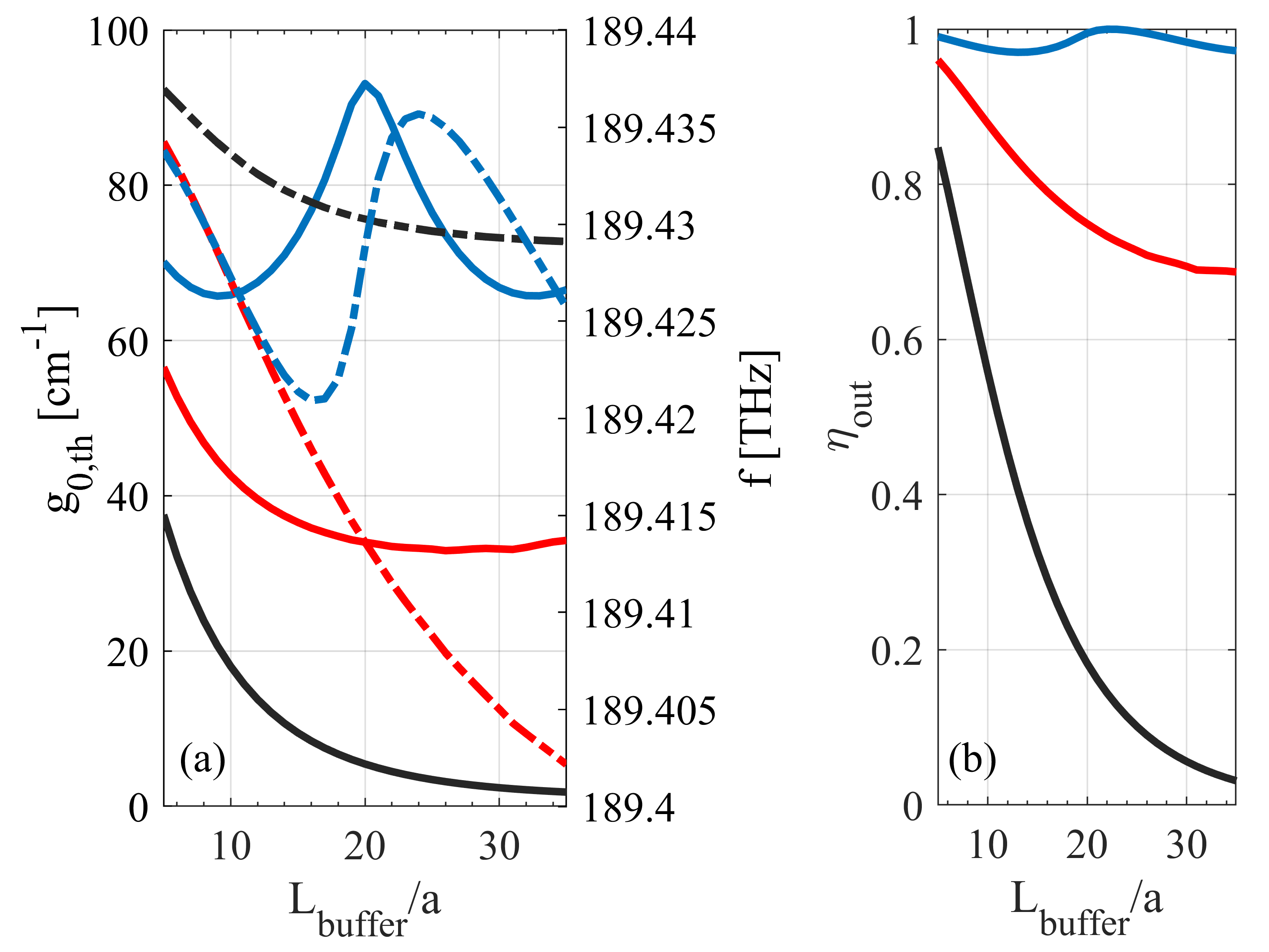}
	\caption{\label{fig:gth_fTHzth_etaOut_TypeABC_DoubleAxes} Threshold gain (solid line) and lasing frequency (dashed line) as a function of the buffer mirror length (a) for type A (red), B (blue) and C (black). Output coupling efficiency (b). The active section has $\Delta n_s = -0.001$ and $L_{\mathrm{active}}=10a$.}
\end{figure}
Fig.~\ref{fig:MirrorsSpectra} shows the reflectivity of the broad-band mirror (green) and that of the buffer mirror in the three different operating conditions: Type A (red), Type B (blue) and Type C (black). In all three cases, the buffer length is fixed to $L_{\mathrm{buffer}} = 15a$. The bullets denote the position of the lasing mode (calculated in the following) in the three cases, with respect to both the broad-band and buffer mirror reflection spectrum. In Type C (black), the buffer refractive index perturbation coincides with that of the broad-band mirror. Therefore, at the lasing frequency, the reflectivity of both mirrors is high (because the lasing mode lies in the stop-band of the buffer mirror) and the laser threshold behaviour is mainly dominated by the mirrors feedback. In Type A, the buffer refractive index perturbation is $\Delta n_s = -0.001$; the lasing frequency is slightly shifted with respect to Type C, but the buffer reflectivity is smaller. In Type B, the buffer refractive index perturbation is $\Delta n_s = -0.0005$; as a result, the band edge of the buffer dispersion relation is in the stopband of the active section. The lasing frequency is practically the same as for type A, but the buffer reflectivity is even smaller. In Type A and B, we can expect the interplay of the feedback in both the active section and the mirrors to play a role in determining the laser threshold.
\begin{figure}
	\centering\includegraphics[width=\linewidth]{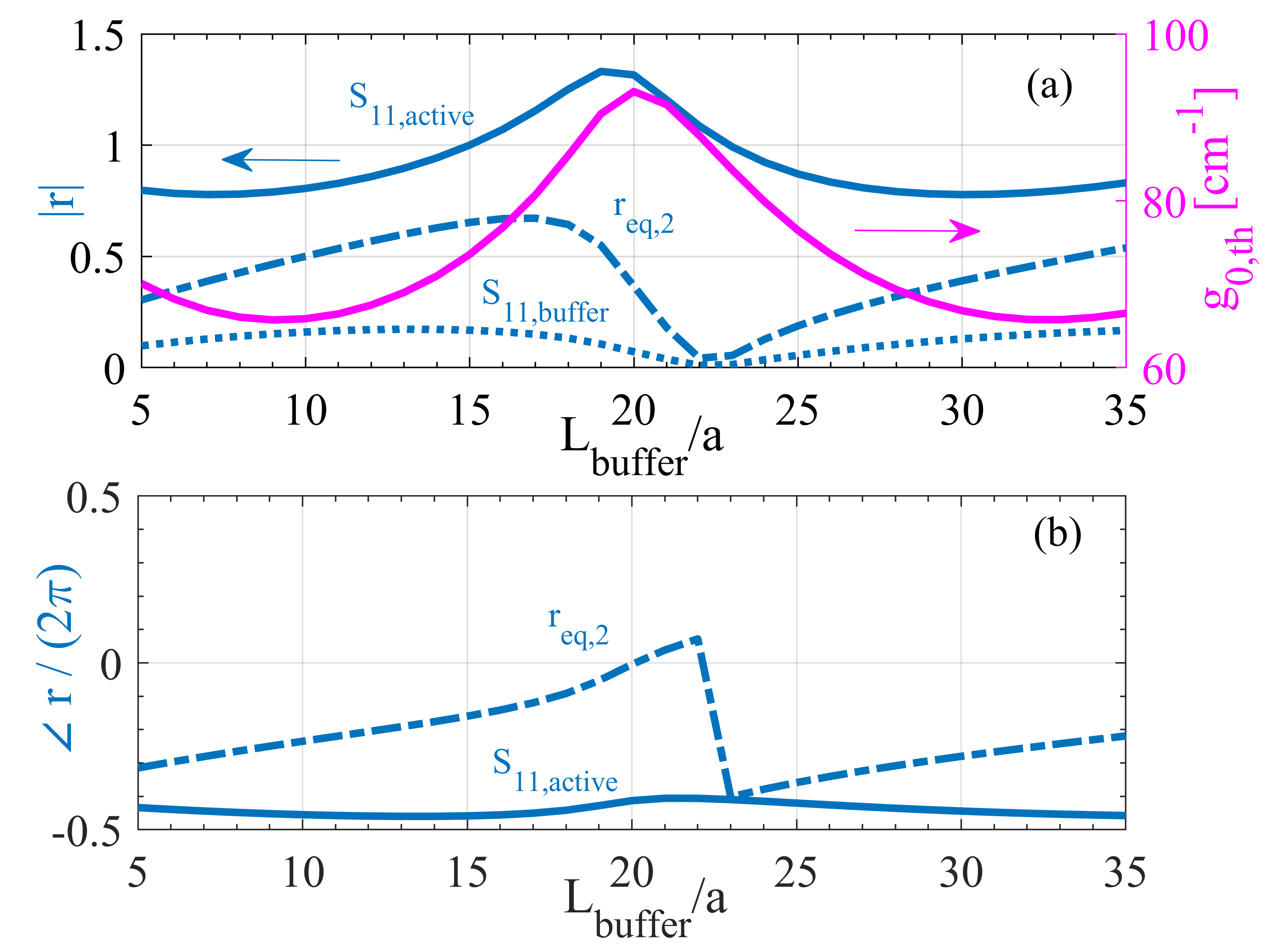}
	\caption{\label{fig:reflTermsMAG_ANGLE_TypeB_separate_WithThresholdGain}\uline{Scattering terms of Eq.~(\ref{eq:reqR}) for Type B configuration of Fig.~\ref{fig:TypeABCcavScheme} evaluated at the lasing frequency and threshold gain. Magnitude of $S_{\mathrm{11,active}}$ (blue, solid line), $r_{\mathrm{eq,2}}$ (blue, dashed line) and $S_{\mathrm{11,buffer}}$ (blue, dotted line); threshold gain (pink line) (a).  Phase of $S_{\mathrm{11,active}}$ (blue, solid line) and $r_{\mathrm{eq,2}}$ (blue, dashed line) (b).}}
\end{figure} 
\begin{figure}
	\centering\includegraphics[width=\linewidth]{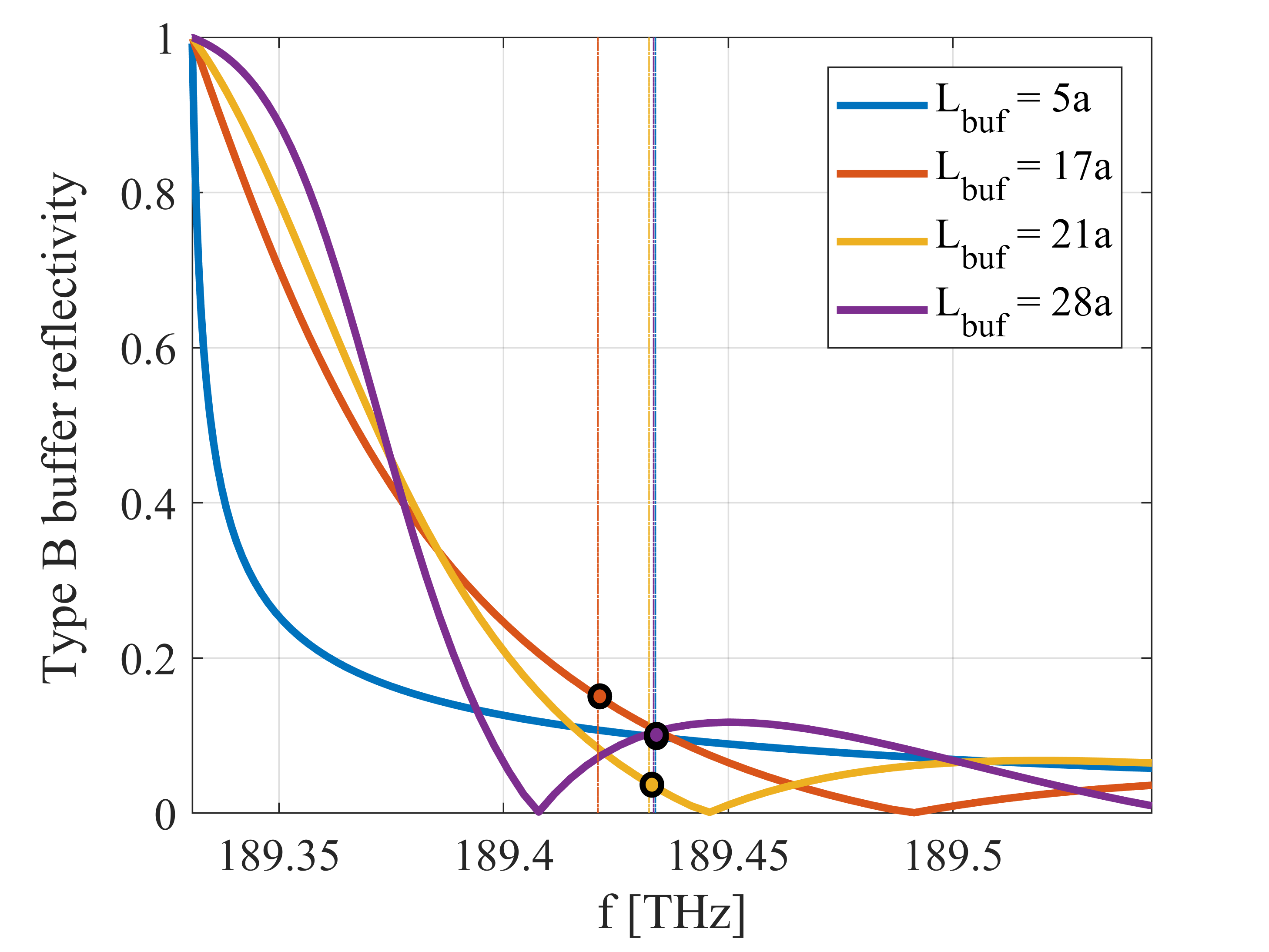}
	\caption{\label{fig:S11mirror_Lbuf5a17a21a28a} Type B buffer mirror reflectivity for different buffer lengths; the bullets denote, for each length, the lasing mode position.}
\end{figure} 
Fig.~\ref{fig:gth_fTHzth_etaOut_TypeABC_DoubleAxes}(a) shows threshold gain (solid curve) and lasing frequency (dashed curve); Fig.~\ref{fig:gth_fTHzth_etaOut_TypeABC_DoubleAxes}(b) reports the output coupling efficiency $\eta_{\mathrm{out}}$, calculated as power transmission through the buffer, as a function of the buffer mirror length. Type C configuration has the smallest threshold gain, which monotonically decreases with increasing buffer length. In this case, the lasing mode is in the stop-band of the buffer mirror, whose reflectivity rapidly increases with increasing number of unit cells; as a result, the power transmission through the buffer rapidly deteriorates with increasing buffer length. This situation is similar to the operating condition of typical LEAP lasers \cite{MatsuoNatPhot2013,MatsuoJSTQE2013}. In Type A configuration, the distributed feedback in the active section provides a back reflection comparable with that of the buffer mirror. In this case, the threshold gain is larger than in type C, but the output coupling efficiency is high and does not significantly deteriorate with increasing buffer length. In Type B configuration, the lasing mode lies outside the stop-band of the buffer mirror; consequently, this configuration exhibits the largest threshold gain, but also the largest $\eta_{\mathrm{out}}$. In this case, the buffer mirror reflectivity is very low and lasing is mainly possible thanks to the active section distributed feedback. Nevertheless, the buffer also slightly contributes to the laser threshold behaviour. The interplay between buffer back reflection and distributed feedback in the active region can be understood by analyzing, with reference to Fig.~\ref{fig:TypeABCcavScheme}, the expression for $r_{\mathrm{eq,R}}$ \uline{given by Eqs.~(\ref{eq:reqR})}. \sout{For the sake of convenience, we denote by $r_{\mathrm{eq,2}}$ the second term on the RHS of Eq.~(\ref{eq:reqR})}. Fig.~\ref{fig:reflTermsMAG_ANGLE_TypeB_separate_WithThresholdGain} shows in blue for Type B the magnitude (a) and phase (b) of the scattering terms, evaluated at the lasing frequency and threshold gain, which contribute to $r_{\mathrm{eq,R}}$; it also reports the corresponding threshold gain (pink line, (a)). Fig.~\ref{fig:S11mirror_Lbuf5a17a21a28a} shows the buffer reflectivity for various buffer lengths; the bullets denote the position of the corresponding lasing mode.
For $L_{\mathrm{buffer}}=5a$, the threshold gain is $g_{\mathrm{0,th}}\simeq \mathrm{70\, cm^{-1}}$ and the lasing frequency $f\simeq\mathrm{189.434\, THz}$; at $L_{\mathrm{buffer}}=21a$, the lasing frequency is practically the same, but the threshold gain is much larger. This is because the mode is close to the zero of the buffer reflection spectrum for $L_{\mathrm{buffer}}=21a$; as a result, although the LG phase condition is satisfied at the same frequency for these two different buffer lengths, $|r_{\mathrm{eq,R}}|$ is smaller in the second case, thus requiring a larger gain for achieving threshold. For $L_{\mathrm{buffer}}=28a$, the lasing frequency is the same again, but the threshold gain practically coincides with that of the structure with $L_{\mathrm{buffer}}=5a$. This is because the mode is now close to the top of the buffer reflection spectrum side-lobe, rather than to the zero (see Fig.~\ref{fig:S11mirror_Lbuf5a17a21a28a}); furthermore, as for $L_{\mathrm{buffer}}=5a$, $S_{\mathrm{11,active}}$ and $r_{\mathrm{eq,2}}$ are nearly in-phase (see Fig.~\ref{fig:reflTermsMAG_ANGLE_TypeB_separate_WithThresholdGain}). For the case of $L_{\mathrm{buffer}}=17a$, although $|r_{\mathrm{eq,2}}|$ is maximum, the threshold gain is not minimum, but rather close to its maximum value. The reason is that $S_{\mathrm{11,active}}$ and $r_{\mathrm{eq,2}}$ are now nearly out-of-phase. \uline{We note that $S_{\mathrm{11,active}}$ and $r_{\mathrm{eq,2}}$ are also nearly out-of-phase for the case $L_{\mathrm{buffer}}=21a$; in this case, however, the magnitude of $r_{\mathrm{eq,2}}$ is too small to significantly affect $r_{\mathrm{eq,R}}$}. As a result of this complex interaction between the active section distributed feedback and the buffer back reflection, the Type B configuration exhibits an optimum number of buffer cells minimizing the threshold gain. These examples prove the great impact of coherent distributed feedback effects in PhC cavities like that in Fig.~\ref{fig:TypeABCcavScheme}, which is similar to a LEAP laser with an in-line coupled waveguide.
  
\begin{figure}
	\centering\includegraphics[width=\linewidth]{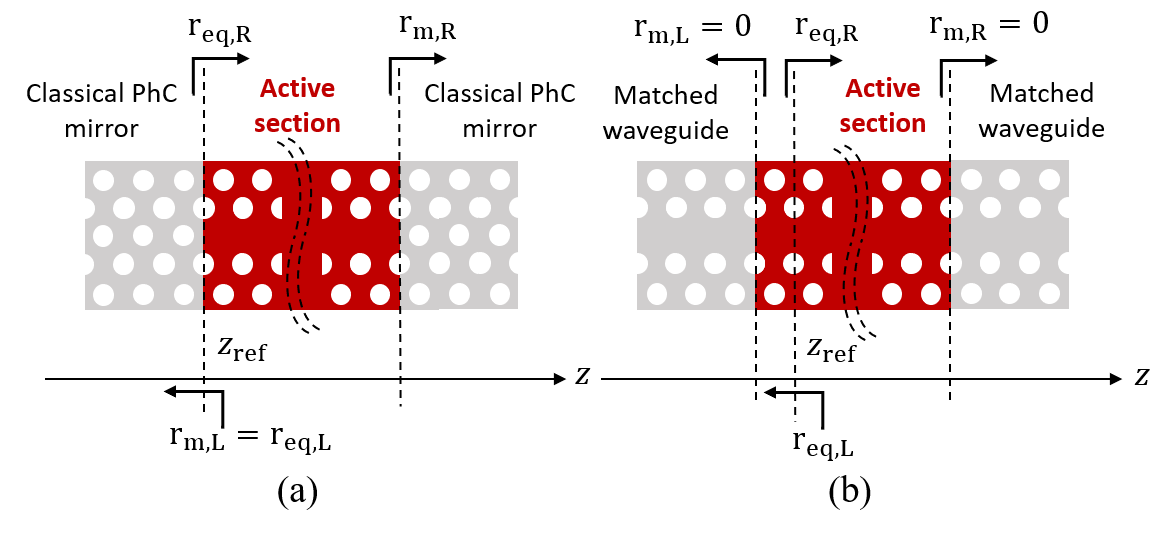}
	\caption{\label{fig:HighZeroFacetReflcavScheme} Typical PhC laser based on a $\mathrm{LN}$ cavity (a). PhC laser cavity consisting of an active section with zero back reflections \uline{for Bloch modes} at the interfaces with the passive output waveguides (b). In both cases, the reference waveguide has $n_s = 3.171$ and the active section $\Delta n_s = -0.001$. \uline{The position of the reference plane to compute the complex loop-gain is denoted by $z_{\mathrm{ref}}$}.}
\end{figure} 
As a second example, we analyze the PhC lasers shown in Fig.~\ref{fig:HighZeroFacetReflcavScheme}(a). This configuration is similar to the optically pumped PhC laser  of \cite{Xue2016}, where the cavity mirrors are classical PhC mirrors. Referring to Fig.~\ref{fig:HighZeroFacetReflcavScheme}(a), $r_{\mathrm{m,R}}$ is the reflectivity that the forward-propagating Bloch mode of the reference, passive waveguide (whose perturbation is accounted for, in the active section, by the coupled-Bloch-mode equations) undergoes when impinging on the right mirror; this mode is reflected back to the active section because it becomes evanescent within the mirror. A similar interpretation holds for $r_{\mathrm{m,L}}$. This reflectivity cannot be computed by our approach, because the classical PhC mirror cannot be viewed as a weak perturbation to a reference waveguide (due to the high refractive index contrast between slab and air holes). However, it has been computed in \cite{Lalanne2008} by modelling the cavity as an effective Fabry-Perot (FP) resonator and by fitting the Q-factor with that obtained through a RCWA approach \cite{LalanneRCWA2007}. For the sake of simplicity and neglecting the impact of disorder, we assume a high, frequency-independent reflectivity $r_{\mathrm{m,L}} = r_{\mathrm{m,R}} = 0.98$, which represents a reasonable approximation \cite{Xue2016,Lalanne2008,Rigal2017}.
\begin{figure}
	\centering\includegraphics[width=\linewidth]{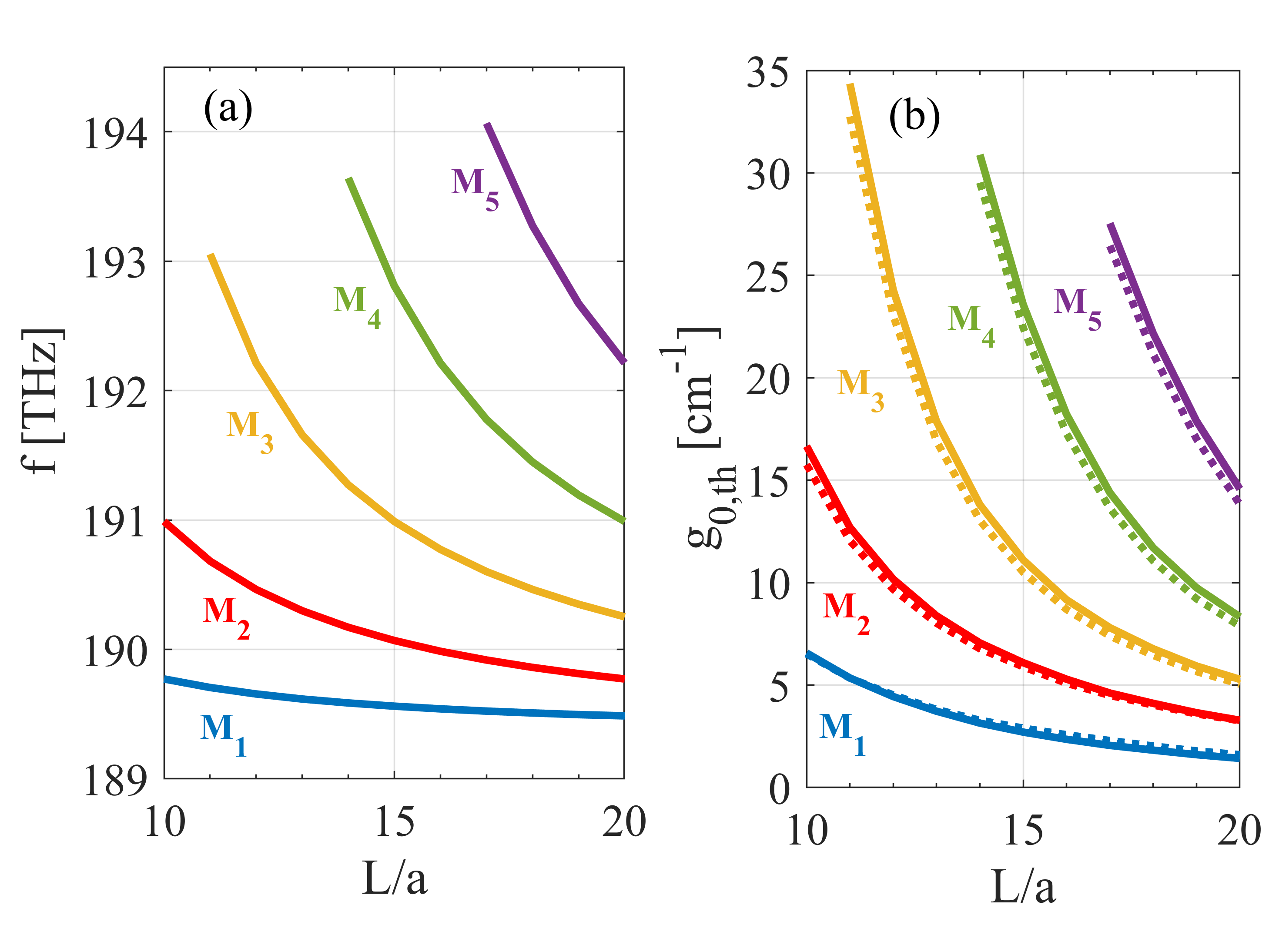}
	\caption{\label{fig:rFacet098_fTHz_g0th} Numerically computed resonant frequencies (a) and threshold gain ((b), solid curve) for the laser in Fig.~\ref{fig:HighZeroFacetReflcavScheme}(a). The threshold gain estimated by the SL-enhanced FP formula is also shown ((b), dotted curve). Each colour corresponds to a different longitudinal resonant mode.}
\end{figure}
\begin{figure}
	\centering\includegraphics[width=\linewidth]{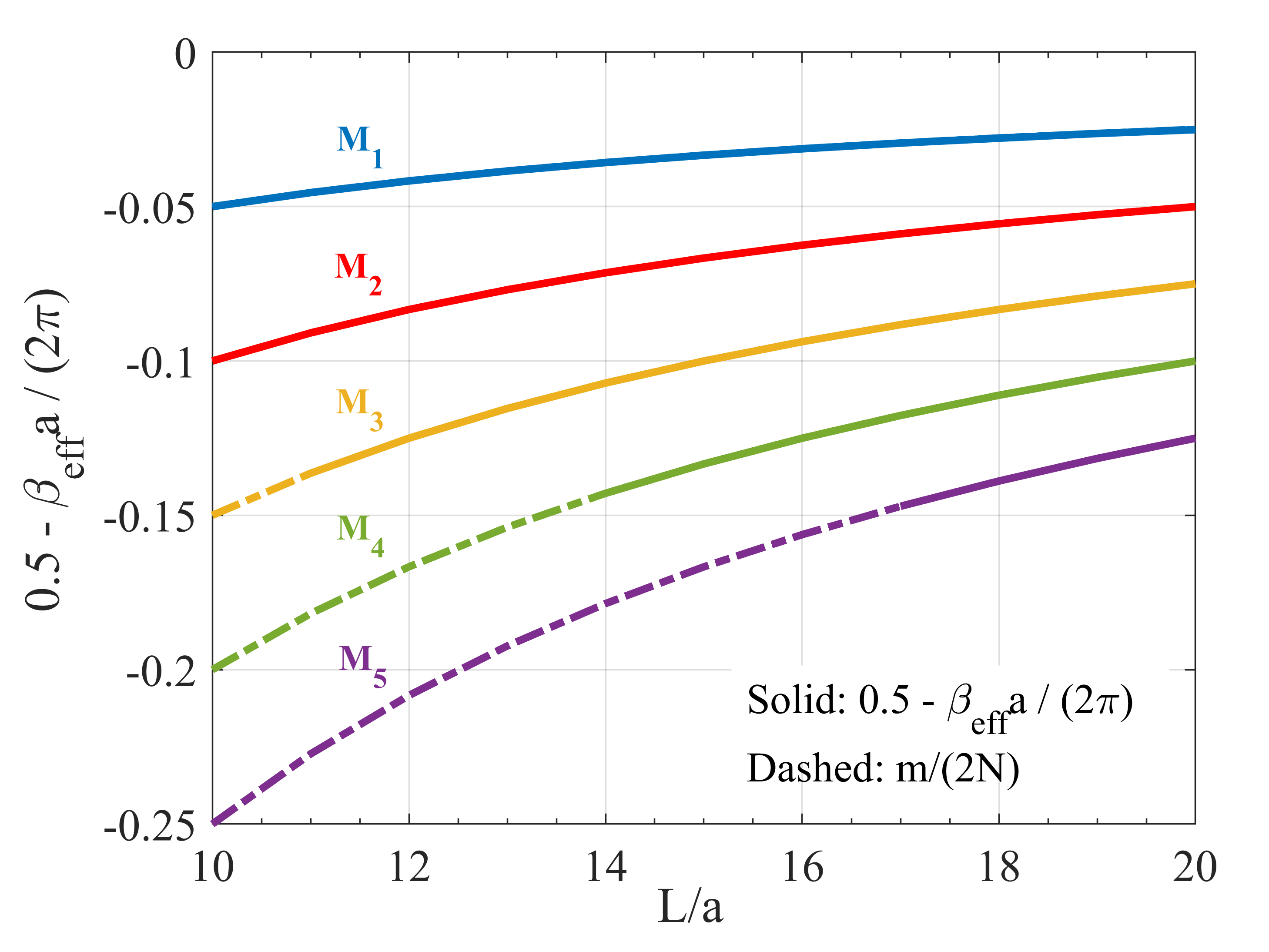}
	\caption{\label{fig:rFacet098_resModeLocations}. Resonant mode condition for the laser in Fig.~\ref{fig:HighZeroFacetReflcavScheme}(a). Each colour corresponds to a different longitudinal resonant mode. $\beta_{\mathrm{eff}}$ is evaluated at the numerically computed resonant frequencies and corresponding threshold gain.}
\end{figure}
To compute the LG, we choose the reference plane at the interface between the active section and the left mirror (see Fig.~\ref{fig:HighZeroFacetReflcavScheme}(a)).
\uline{As a result, $r_{\mathrm{eq,L}}$ is equal to $r_{\mathrm{m,L}}$ and $r_{\mathrm{eq,R}}$ is obtained from Eq.~(\ref{eq:reqR}) by replacing $S_{\mathrm{11,buffer}}$ with $r_{\mathrm{m,R}}$}. The laser threshold behaviour is summarized in Fig.~\ref{fig:rFacet098_fTHz_g0th}, showing the numerically computed resonant frequencies (a) and threshold gain ((b), solid curve); each colour corresponds to a different longitudinal resonant mode. In the existing literature, similar cavities have been studied in \cite{Okano2010} through FDTD simulations. In \cite{Okano2010}, it has been shown that the resonant modes of a passive $\mathrm{LN}$ cavity correspond to the fullfillment of the condition
\begin{equation}
\label{eq:ResCond}
\left(0.5 - k_za/2\pi\right) = m/2N
\end{equation}
with $N = L/a$ being the number of unit cells, $m$ the mode order and $k_z$ the dispersion relation of the passive PhC LDWG of the cavity. For this reason, we have evaluated the quantity $0.5 - \beta_{\mathrm{eff}}a/2\pi$ at the numerically computed resonant frequencies and corresponding threshold gain of Fig.~\ref{fig:rFacet098_fTHz_g0th}; it is shown as the solid curve in Fig.~\ref{fig:rFacet098_resModeLocations}. We also note that 
the quantity $0.5 - \beta_{\mathrm{eff}}a/2\pi$ \sout{exactly} \uline{practically} coincides, at a given cavity length, with $m/2N$ (dashed curve in Fig.~\ref{fig:rFacet098_resModeLocations}). Since the required threshold gain is low, the gain-induced distributed feedback is negligible. As a consequence, the gain does not impact on the position of the resonant modes. As the cavity length increases, the modes move towards the SL region along the dispersion relation of the umpumped waveguide. This \uline{effect} is consistent with experimental \cite{Xue2016} and numerical \cite{Cartar2017} trends and is \sout{just a FP effect} \uline{independent of the perturbation-induced distributed feedback}. In fact, by setting $\Delta n_s=0$ and $\kappa_{12,q=1} = \kappa_{21,q=-1} = 0$ in Eqs.~(\ref{eq:Tmatrix}), Eq.~\ref{eq:ResCond} can be easily obtained. The threshold gain reported in Fig.~\ref{fig:rFacet098_fTHz_g0th}(b) is compared with the expression $\left[1/\left(S\: L_{\mathrm{active}}\right)\right]\ln\left[1/\left(r_{\mathrm{m,L}}r_{\mathrm{m,R}}\right) \right]$ (dotted curve in Fig.~\ref{fig:rFacet098_fTHz_g0th}(b)) \cite{Xue2016}, with $S = n_{g}/n_s$ being the slow-down factor, evaluated at the resonant frequencies, and $L_{\mathrm{active}}$ the cavity length; again, the group index is that of the umpumped waveguide. \uline{This expression resembles that of a standard FP laser, but the threshold gain is scaled down by the slow-down factor}. Since the gain is low, $n_g$ is not reduced and the SL enhancement of the gain is not limited. On the basis of these considerations, we conclude that the laser with classical PhC mirrors modelled as high-reflectivity, frequency-independent reflectors behaves as a SL-enhanced FP laser \uline{for Bloch modes}.

\begin{figure}
	\centering\includegraphics[width=\linewidth]{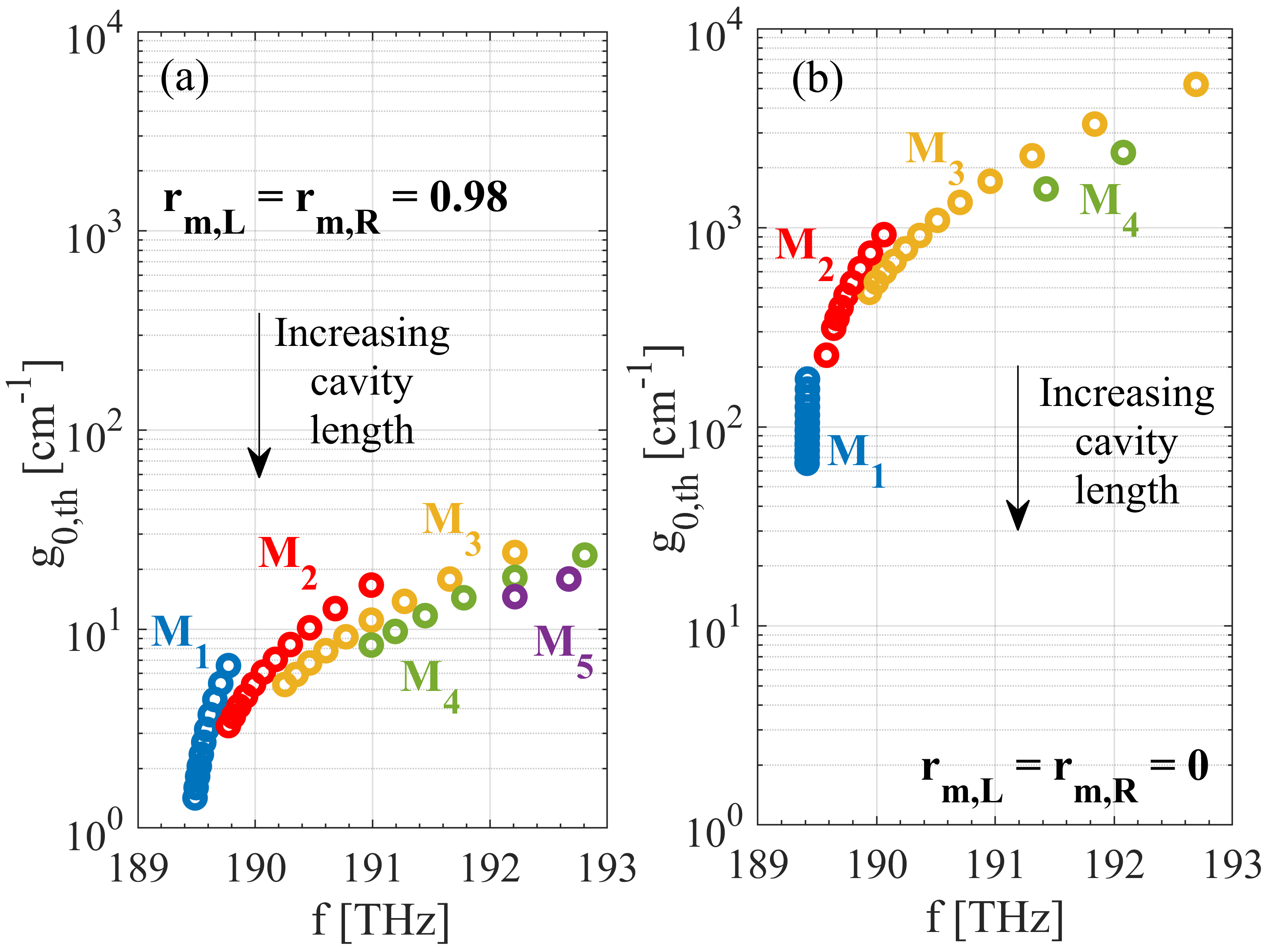}
	\caption{\label{fig:rFacet098Zero_g0th_fth_comparison_withMatlab} Threshold gain and corresponding resonant frequency for the laser in Fig.~\ref{fig:HighZeroFacetReflcavScheme}(a) (a) and Fig.~\ref{fig:HighZeroFacetReflcavScheme}(b) (b) as a function of cavity length. Each colour corresponds to a different longitudinal resonant mode. The cavity length ranges from $10a$ to $20a$.}
\end{figure} 
\begin{figure}
	\centering\includegraphics[width=\linewidth]{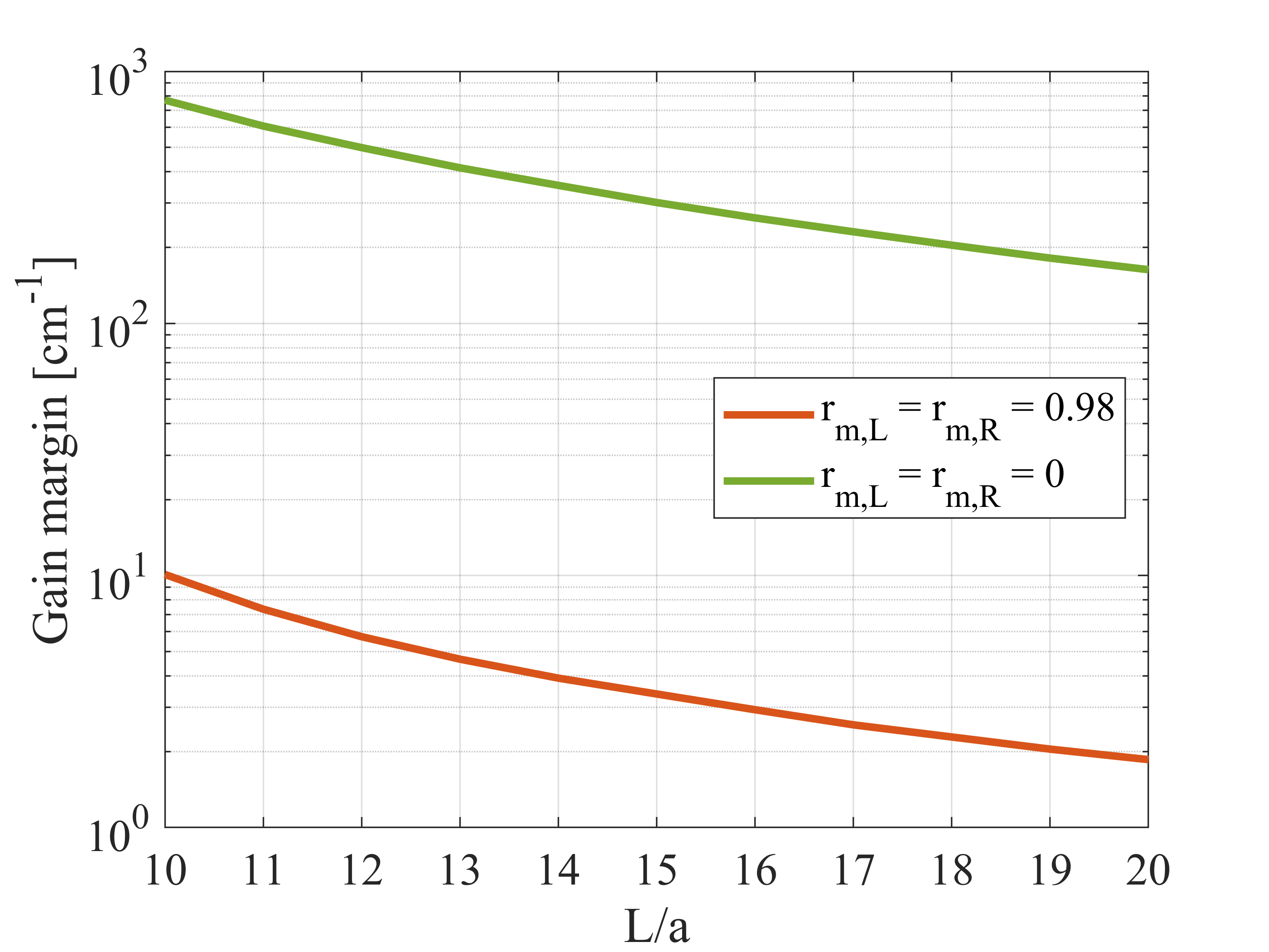}
	\caption{\label{fig:rFacet098Zero_GainMargin_NaEVEN_ODD} Difference between $\mathrm{M_2}$ and $\mathrm{M_1}$ threshold gain for the laser in Fig.~\ref{fig:HighZeroFacetReflcavScheme}(a) (red) and (b) (green) as a function of the cavity length.}
\end{figure}
As a last example, we focus on the structure in Fig.~\ref{fig:HighZeroFacetReflcavScheme}(b), which consists of an active section with zero back reflections \uline{for \textit{Bloch} modes} at the interfaces with the passive output waveguides. \sout{We note that this condition can be practically implemented by making the active section refractive index (or width) sufficiently smaller than that of the output waveguides, such that the cavity resonant modes are in the propagation band of the output waveguides. This implies that, differently from the typical
LEAP laser implementation, the passive sections do not
act as mirrors}. \uline{The practical implementation of this \textit{matching} condition is outside the scope of this paper; various promising solutions are reported in the existing literature \cite{DUTTAreview,Hugonin2007}.} We show for the first time, to the best of our knowledge, that this type of structure can ideed achieve lasing with reasonable threshold gain. \sout{The output waveguides coincide with} The reference waveguide \sout{having} has $n_s = 3.171$, while the active section has $\Delta n_s = -0.001$ and $g_0>0$. \uline{Similarly to the second example, here a given $\Delta n_s$ is only considered in order to shift away the active section dispersion relation from the critical point $k_z = \pi/a$ and investigate the laser operation also at frequencies near the band edge (as explained in the end of section II).} To compute the LG, we choose a reference plane within the active region, at the interface between any two unit cells; $N_R$ unit cells are located on the right of the reference plane and $N_L$ on the left, with $N = N_L + N_R$ being the total number of unit cells in the whole active region. \sout{Therefore, to obtain $r_{\mathrm{eq,R}}$, the transmission matrix of an active section of $N_R$ unit cells is computed by Eqs.~(\ref{eq:FrobeniusTheorem}) and turned into a scattering matrix; the $\mathrm{S_{11}}$ parameter of this scattering matrix is $r_{\mathrm{eq,R}}$. Similarly, $r_{\mathrm{eq,L}}$ is computed as the $S_{22}$ parameter of the scattering matrix corresponding to the other $N_L$ cells.} \uline{Therefore, $r_{\mathrm{eq,R}}$ is given by the $S_{11}$ parameter of Eqs.~(\ref{eq:Smatrix}) with $N = N_R$; similarly, $r_{\mathrm{eq,L}}$ is computed as the $S_{22}$ parameter of Eqs.~(\ref{eq:Smatrix}) with $N = N_L$.} The threshold gain and corresponding resonant frequencies of this laser are shown in Fig.~\ref{fig:rFacet098Zero_g0th_fth_comparison_withMatlab}(b) and compared with those of the laser with classical PhC mirrors (Fig.~\ref{fig:rFacet098Zero_g0th_fth_comparison_withMatlab}(a)). The cavity length ranges from $10a$ to $20a$ and each colour corresponds to a different longitudinal resonant mode. As a first remark, we note that the threshold gain is considerably larger for the laser with zero back reflections; however, the threshold gain of the lasing mode ($\mathrm{M_1}$) turns out to be reasonable, being of the same order of magnitude as observed for the Type B configuration in Fig.~\ref{fig:TypeABCcavScheme}. Secondly, the lasing mode frequency is essentially independent of the cavity length and it is located close to the band edge of the active section dispersion relation with $g_0 = 0$. On the contrary, the higher-order modes shift towards the SL region as the cavity length increases. This is somehow similar to what occurs in purely gain-coupled DFB lasers, whose lasing frequency is independent of the cavity length and exactly located at the Bragg frequency \cite{KogelnikShank}. Finally, we note that the laser in Fig.~\ref{fig:HighZeroFacetReflcavScheme}(b) exhibits a much better spectral selectivity as compared to a laser with classical PhC mirrors. This is illustrated in Fig.~\ref{fig:rFacet098Zero_GainMargin_NaEVEN_ODD}, comparing the gain margin (defined as the difference between $\mathrm{M_2}$ and $\mathrm{M_1}$ threshold gain)  for the lasers in Fig.~\ref{fig:HighZeroFacetReflcavScheme} as a function of cavity length.

\section{Conclusions}

By starting from a set of two coupled-Bloch-mode equations \cite{Chen2015}, we have derived a simple, closed-form expression for the unit cell transmission matrix of a PhC LDWG with a generally complex refractive index perturbation as compared to a reference waveguide. This allows for a simple and numerically efficient analysis of active PhC LDWGs and lasers based on this type of waveguide, such as LEAP lasers.

In particular, we have derived the expression of the coupling coefficients and explained that the magnitude of the cross-coupling is always comparable to that of self-coupling; this is due to the non-negligible longitudinal component of TE-like Bloch modes in PhC LDWGs. We have shown that our approach can correctly reproduce the formation of a stop-band for Bloch modes as a consequence of a purely real refractive index perturbation. We have further validated it by computing the group index and gain enhancement factor of an active PhC waveguide; consistently with the rigorous, non-perturbative approach of \cite{Gric2012}, we have shown that the maximum attainable SL gain enhancement is limited by the gain itself.          

We have then applied our coupled-Bloch-mode approach to analyze the threshold condition of three types of PhC laser cavities. The first cavity is conceptually similar to that characterized in \cite{Shinya}. Depending on the buffer refractive index perturbation, we have identified, consistently with \cite{Shinya}, three different operating regimes, thus proving the great impact of coherent distributed feedback effects in this type of PhC cavity. The second cavity is the one characterized in \cite{Xue2016}. By neglecting the impact of fabrication disorder and modelling the classical PhC mirrors as standard reflectors with a high, frequency-independent reflectivity, we have shown that the gain-induced distributed feedback is negligible in this type of cavity, which simply behaves as a SL-enhanced FP laser \uline{for Bloch modes}. As a last example, we have analyzed a structure consisting of an active section bounded on either side by passive waveguides, which are assumed to \sout{provide zero back reflections} \uline{be matched for the reference waveguide Bloch modes}. This means that this configuration is different from the typical LEAP laser implementation. Interestingly, we have shown that this cavity can lase with reasonable threshold gain, with lasing only sustained by the active region distributed feedback. 

In conclusion, we have presented an effective approach that will be useful to provide insights on the characteristics of PhC lasers and might be also extended to study the laser dynamics of these structures.   

\section{Acknowledgements}

M.S. wishes to express his gratitude to Prof. I. Montrosset and Prof. R. Orta for fruitful discussions on CMT and Bloch modes. J. Mørk acknowledges support for the Villum Foundation through the NATEC Centre of Excellence (grant 8692). 

\bibliographystyle{IEEEtran}
\bibliography{JSTQE_MS_ArXiV}

\begin{thebibliography}{10}
\providecommand{\url}[1]{#1}
\csname url@samestyle\endcsname
\providecommand{\newblock}{\relax}
\providecommand{\bibinfo}[2]{#2}
\providecommand{\BIBentrySTDinterwordspacing}{\spaceskip=0pt\relax}
\providecommand{\BIBentryALTinterwordstretchfactor}{4}
\providecommand{\BIBentryALTinterwordspacing}{\spaceskip=\fontdimen2\font plus
\BIBentryALTinterwordstretchfactor\fontdimen3\font minus
  \fontdimen4\font\relax}
\providecommand{\BIBforeignlanguage}[2]{{%
\expandafter\ifx\csname l@#1\endcsname\relax
\typeout{** WARNING: IEEEtran.bst: No hyphenation pattern has been}%
\typeout{** loaded for the language `#1'. Using the pattern for}%
\typeout{** the default language instead.}%
\else
\language=\csname l@#1\endcsname
\fi
#2}}
\providecommand{\BIBdecl}{\relax}
\BIBdecl

\bibitem{Baba2008}
T.~Baba, ``Slow light in photonic crystals,'' \emph{Nat. Photon.}, vol.~2,
  no.~8, pp. 465--473, 2008.

\bibitem{ek2014a}
S.~Ek, P.~L. Hansen, Y.~Chen, E.~Semenova, K.~Yvind, and J.~Mørk,
  ``Slow-light-enhanced gain in active photonic crystal waveguides,''
  \emph{Nat. Commun.}, vol.~5, 2014.

\bibitem{MatsuoReview}
S.~Matsuo and T.~Kakitsuka, ``Low-operating-energy directly modulated lasers
  for short-distance optical interconnects,'' \emph{Adv. Opt. Photon.},
  vol.~10, no.~3, pp. 567--643, Sep. 2018.

\bibitem{Okano2010}
M.~Okano, T.~Yamada, J.~Sugisaka, N.~Yamamoto, M.~Itoh, T.~Sugaya, K.~Komori,
  and M.~Mori, ``Analysis of two-dimensional photonic crystal {L}-type cavities
  with low-refractive-index material cladding,'' \emph{J. Opt.}, vol.~12,
  no.~7, p. 075101, Jun. 2010.

\bibitem{Nomura2006}
M.~Nomura, S.~Iwamoto, K.~Watanabe, N.~Kumagai, Y.~Nakata, S.~Ishida, and
  Y.~Arakawa, ``Room temperature continuous-wave lasing in photonic crystal
  nanocavity,'' \emph{Opt. Express}, vol.~14, no.~13, pp. 6308--6315, Jun.
  2006.

\bibitem{Matsuo2012}
S.~Matsuo, K.~Takeda, T.~Sato, M.~Notomi, A.~Shinya, K.~Nozaki, H.~Taniyama,
  K.~Hasebe, and T.~Kakitsuka, ``Room-temperature continuous-wave operation of
  lateral current injection wavelength-scale embedded active-region
  photonic-crystal laser,'' \emph{Opt. Express}, vol.~20, no.~4, pp.
  3773--3780, Feb. 2012.

\bibitem{Xue2015}
W.~Xue, L.~Ottaviano, Y.~Chen, E.~Semenova, Y.~Yu, A.~Lupi, J.~M\o{}rk, and
  K.~Yvind, ``Thermal analysis of line-defect photonic crystal lasers,''
  \emph{Opt. Express}, vol.~23, pp. 18\,277--18\,287, Jul. 2015.

\bibitem{Xue2016}
W.~Xue, Y.~Yu, L.~Ottaviano, Y.~Chen, E.~Semenova, K.~Yvind, and J.~M\o{}rk,
  ``Threshold characteristics of slow-light photonic crystal lasers,''
  \emph{Phys. Rev. Lett.}, vol. 116, no.~6, p. 063901, Feb. 2016.

\bibitem{MatsuoNatPhot2013}
K.~Takeda, T.~Sato, A.~Shinya, K.~Nozaki, W.~Kobayashi, H.~Taniyama, M.~Notomi,
  K.~Hasebe, T.~Kakitsuka, and S.~Matsuo, ``Few-{fJ/bit} data transmissions
  using directly modulated lambda-scale embedded active region photonic-crystal
  lasers,'' \emph{Nat. Photon.}, vol.~7, no.~7, p. 569, 2013.

\bibitem{MatsuoJSTQE2013}
S.~Matsuo, T.~Sato, K.~Takeda, A.~Shinya, K.~Nozaki, H.~Taniyama, M.~Notomi,
  K.~Hasebe, and T.~Kakitsuka, ``Ultralow operating energy electrically driven
  photonic crystal lasers,'' \emph{IEEE J. Sel. Top. Quantum Electron.},
  vol.~19, no.~4, pp. 4\,900\,311--4\,900\,311, Jul. 2013.

\bibitem{Shinya}
A.~Shinya, T.~Sato, K.~Takeda, K.~Nozaki, E.~Kuramochi, T.~Kakitsuka,
  H.~Taniyama, T.~Fujii, S.~Matsuo, and M.~Notomi, ``Single-mode lasing of
  {L}ambda-scale embedded active-region photonic-crystal ({LEAP}) laser with
  in-line coupled waveguide,'' in \emph{2013 IEEE Photonics Conference}, Sep.
  2013, pp. 448--449.

\bibitem{Okano2009}
M.~Okano, T.~Yamada, J.~Sugisaka, N.~Yamamoto, M.~Itoh, T.~Sugaya, K.~Komori,
  and M.~Mori, ``Design of two-dimensional photonic crystal nanocavities with
  low-refractive-index material cladding,'' \emph{J. Opt.}, vol.~12, p. 015108,
  Nov. 2009.

\bibitem{LalanneRCWA2005}
J.~P. Hugonin and P.~Lalanne, ``Perfectly matched layers as nonlinear
  coordinate transforms: a generalized formalization,'' \emph{J. Opt. Soc. Am.
  A}, vol.~22, no.~9, pp. 1844--1849, Sep. 2005.

\bibitem{LalanneRCWA2007}
G.~Lecamp, J.~P. Hugonin, and P.~Lalanne, ``Theoretical and computational
  concepts for periodic optical waveguides,'' \emph{Opt. Express}, vol.~15,
  no.~18, pp. 11\,042--11\,060, Sep. 2007.

\bibitem{SongRCWA}
A.~Y. Song, A.~R.~K. Kalapala, W.~Zhou, and S.~Fan, ``First-principles
  simulation of photonic crystal surface-emitting lasers using rigorous coupled
  wave analysis,'' \emph{Appl. Phys. Lett.}, vol. 113, no.~4, p. 041106, 2018.

\bibitem{Yariv1973}
A.~Yariv, ``Coupled-mode theory for guided-wave optics,'' \emph{IEEE J. Quantum
  Electron.}, vol.~9, no.~9, pp. 919--933, Sep. 1973.

\bibitem{Marcuse_Book1974}
D.~Marcuse, \emph{Theory of Dielectric Optical Waveguides}, 2nd~ed.\hskip 1em
  plus 0.5em minus 0.4em\relax Academic Press, 1991.

\bibitem{KogelnikShank}
H.~Kogelnik and C.~V. Shank, ``Coupled‐wave theory of {D}istributed
  {F}eedback {L}asers,'' \emph{J. Appl. Phys.}, vol.~43, no.~5, pp. 2327--2335,
  1972.

\bibitem{Patterson2010}
M.~Patterson and S.~Hughes, ``Theory of disorder-induced coherent scattering
  and light localization in slow-light photonic crystal waveguides,'' \emph{J.
  Opt.}, vol.~12, no.~10, p. 104013, 2010.

\bibitem{MichaelisPR}
D.~Michaelis, U.~Peschel, C.~W\"achter, and A.~Br\"auer, ``Reciprocity theorem
  and perturbation theory for photonic crystal waveguides,'' \emph{Phys. Rev.
  E}, vol.~68, no.~6, p. 065601, Dec. 2003.

\bibitem{Chen2015}
Y.~Chen, J.~R. de~Lasson, N.~Gregersen, and J.~M\o{}rk, ``Impact of slow-light
  enhancement on optical propagation in active semiconductor photonic-crystal
  waveguides,'' \emph{Phys. Rev. A}, vol.~92, no.~5, p. 053839, Nov. 2015.

\bibitem{Gric2012}
J.~Grgi\ifmmode~\acute{c}\else \'{c}\fi{}, J.~R. Ott, F.~Wang, O.~Sigmund,
  A.-P. Jauho, J.~M\o{}rk, and N.~A. Mortensen, ``Fundamental limitations to
  gain enhancement in periodic media and waveguides,'' \emph{Phys. Rev. Lett.},
  vol. 108, p. 183903, May 2012.

\bibitem{DeRossi2009}
Q.~V. Tran, S.~Combrié, P.~Colman, and A.~De~Rossi, ``Photonic crystal
  membrane waveguides with low insertion losses,'' \emph{Applied Physics
  Letters}, vol.~95, no.~6, p. 061105, 2009.

\bibitem{MPBpaper}
S.~G. Johnson and J.~D. Joannopoulos, ``Block-iterative frequency-domain
  methods for {M}axwell's equations in a planewave basis,'' \emph{Opt.
  Express}, vol.~8, no.~3, pp. 173--190, Jan. 2001.

\bibitem{skorobogatiy_yang_2008}
M.~Skorobogatiy and J.~Yang, \emph{Fundamentals of Photonic Crystal
  Guiding}.\hskip 1em plus 0.5em minus 0.4em\relax Cambridge University Press,
  2008.

\bibitem{Frobenius}
L.~A. Pipes and S.~A. Hovanesian, \emph{Matrix-computer methods in
  engineering}.\hskip 1em plus 0.5em minus 0.4em\relax John Wiley \& Sons,
  1969.

\bibitem{Coldren}
L.~A. Coldren, S.~W. Corzine, and M.~L. Ma\v{s}anovi\'{c}, \emph{Diode Lasers
  and Photonic Integrated Circuits}.\hskip 1em plus 0.5em minus 0.4em\relax
  John Wiley \& Sons, 2012.

\bibitem{Matsuo2010}
S.~Matsuo, A.~Shinya, T.~Kakitsuka, K.~Nozaki, T.~Segawa, T.~Sato,
  Y.~Kawaguchi, and M.~Notomi, ``High-speed ultracompact buried heterostructure
  photonic-crystal laser with 13 {fJ} of energy consumed per bit transmitted,''
  \emph{Nat. Photon.}, vol.~4, Aug. 2010.

\bibitem{Notomi2001}
M.~Notomi, K.~Yamada, A.~Shinya, J.~Takahashi, C.~Takahashi, and I.~Yokohama,
  ``Extremely large group-velocity dispersion of line-defect waveguides in
  photonic crystal slabs,'' \emph{Phys. Rev. Lett.}, vol.~87, Jan. 2001.

\bibitem{Lalanne2008}
P.~Lalanne, C.~Sauvan, and J.~P. Hugonin, ``Photon confinement in photonic
  crystal nanocavities,'' \emph{Laser Photonics Rev.}, vol.~2, pp. 514--526,
  Dec. 2008.

\bibitem{Rigal2017}
B.~Rigal, K.~Joanesarson, A.~Lyasota, C.~Jarlov, B.~Dwir, A.~Rudra, I.~Kulkova,
  and E.~Kapon, ``Propagation losses in photonic crystal waveguides: effects of
  band tail absorption and waveguide dispersion,'' \emph{Opt. Express},
  vol.~25, p. 28908, Nov. 2017.

\bibitem{Cartar2017}
W.~Cartar, J.~M\o{}rk, and S.~Hughes, ``Self-consistent {M}axwell-{B}loch model
  of quantum-dot photonic-crystal-cavity lasers,'' \emph{Phys. Rev. A},
  vol.~96, Aug. 2017.

\bibitem{DUTTAreview}
H.~S. Dutta, A.~K. Goyal, V.~Srivastava, and S.~Pal, ``Coupling light in
  photonic crystal waveguides: A review,'' \emph{Photonics and Nanostructures -
  Fundamentals and Applications}, vol.~20, pp. 41 -- 58, 2016.

\bibitem{Hugonin2007}
J.~P. Hugonin, P.~Lalanne, T.~P. White, and T.~F. Krauss, ``Coupling into
  slow-mode photonic crystal waveguides,'' \emph{Opt. Lett.}, vol.~32, no.~18,
  pp. 2638--2640, Sep 2007.

\end{thebibliography}

\begin{IEEEbiographynophoto}{Marco Saldutti}
	received the B.Sc. degree in Electronic Engineering from the University of Salerno, Salerno, Italy, in 2014 and the M.Sc. degree in Nanotechnologies for ICTs in 2017 from the Politecnico di Torino, Turin, Italy, where he is currently pursuing the Ph.D. degree. He carried out his M.Sc. project on active photonic crystal structures at the Technical University of Denmark - DTU Fotonik, Lyngby, as a visiting student. His current research interests include modelling of active photonic crystal waveguides and lasers and sources for silicon photonics integrated circuits. 
\end{IEEEbiographynophoto}

\begin{IEEEbiographynophoto}{Paolo Bardella} 
    (M'17) received the M.Sc. degree in Electronic Engineering and	the Ph.D. degree in Electronic and Communications Engineering from the Politecnico di Torino, Turin, Italy, in 2001 and 2006, respectively. He is currently an Associate Professor at the Department of Electronics and Telecommunication of Politecnico di Torino, where he works on the modelling of edge emitting semiconductor lasers based on quantum well and quantum dot active materials, and on the design of passive components for silicon photonics. 
\end{IEEEbiographynophoto}

\begin{IEEEbiographynophoto}{Jesper Mørk} 
	received the M.Sc., Ph.D., and Dr. Techn. degrees from the Technical University of Denmark (DTU), Lyngby, in 1986, 1988, and 2003, respectively. Since 2002 he is a Professor in semiconductor photonics and since 2008 he is heading the Nanophotonics Section at DTU Fotonik, Technical University of Denmark. He is the author of more than 245 papers in refereed journals and around 350 contributions to international conferences, including more than 80 invited talks. His current research interests include semiconductor quantum photonics, photonic crystal structures, slow light, nano- and micro-cavity lasers and integrated photonics. Jesper Mørk is a Fellow of OSA and serves as Associate Editor of Optica. 
\end{IEEEbiographynophoto}

\begin{IEEEbiographynophoto}{Mariangela Gioannini}
	received the Ph.D. degree in Electronic and Communications Engineering from Politecnico di Torino in 2002. She is currently an Associate Professor at the Department of Electronics and Telecommunication of Politecnico di Torino. Her research interests are in the field of photonic devices, where she is working on the development of simulation tools and models to study and design semiconductor light sources with particular focus on: quantum dot lasers and SLEDs, nano-lasers and laser integration in silicon photonics platform. She has also been involved in research projects on the design of third generation solar cells.
\end{IEEEbiographynophoto}

\vfill
\end{document}